\documentclass[aps,prb,twocolumn,amsmath,amssymb,superscriptaddress,floatfix]{revtex4-2}
\pdfoutput=1

\usepackage{mathrsfs}
\usepackage{amsmath}
\usepackage{amsfonts}
\usepackage{float}
\usepackage{graphicx} % Include figure files
\usepackage{bm} % bold math
\usepackage{xcolor}
\usepackage{comment}
\usepackage{mathtools}
\usepackage{braket}
\begin{document}
\title{Quantum spin fluctuations in dynamical quantum phase transitions}
\author{ Cheuk Yiu Wong }
\affiliation{ \textit Department of Physics, City University of Hong Kong, Kowloon, Hong Kong}
\author{Hadi Cheraghi}
\affiliation{\textit Institute of Physics, Maria Curie-Sk\l odowska University, 20-031 Lublin, Poland}
\author{ Wing Chi Yu }
\email{wingcyu@cityu.edu.hk}
\affiliation{ \textit Department of Physics, City University of Hong Kong, Kowloon, Hong Kong}

\date{\today}

\begin{abstract}
%	Dynamical quantum phase transitions (DQPTs) have long been studied in their relation to quantum fluctuations in critical regimes. These fluctuations can be quantified as the degree of spin-squeezing in spin models, where one of two non-commutative observables breaks the standard quantum limit of measurement by minimizing its uncertainty. Given the recent approach of the possibility to probe DQPTs by spin-squeezing in bosonic model, we combine the Loschmidt amplitude, which detects DQPTs, and the spin-squeezing parameter (SSP), the quantification of spin-squeezing, to study the spin dynamics of a fermionic integrable model. We analytically show the extremal, mostly maximal, value of SSP near DQPTs when quenched the system between different phases. These critical phenomena further unveil the spin correlations during DQPTs, for which the highest contribution aligns with the preferred interaction in the postquenched phase. We also demonstrate the time-evolution of SSP differs for various quench types. Our study provides a further understanding of the relation between DQPTs and quantum fluctuations and suggests a phenomenal view further supporting the notion of phase transitions in dynamic regime.
Quantum phase transitions have long been studied in their relation to quantum fluctuations. These fluctuations can be quantified as the degree of spin squeezing in spin models, where one of the two non-commutative observables breaks the standard quantum limit of measurement by minimizing its uncertainty. However, the understanding of their role in dynamical quantum phase transitions (DQPTs) is still incomplete.  In this work, we combine the Loschmidt amplitude, which detects DQPTs, and the spin-squeezing parameter (SSP), the quantification of spin squeezing, to study the spin dynamics in a quenched interacting spin model around DQPT. We show that the extremal, mostly maximal, of SSP occurs near DQPTs when the system is quenched between different phases. These phenomena further unveil the spin correlations during DQPTs, for which the highest contribution aligns with the preferred direction of spin interactions in the post-quenched phase. We also demonstrate the time evolution of SSP differs for various quench scenarios. These findings provide us with physical insights into the dynamics of quantum fluctuations around DQPTs and their relation to the equilibrium phase diagrams.
\end{abstract}

\maketitle

\section{Introduction}

Dynamical quantum phase transitions (DQPTs) have been serving as a theoretical framework on far-from-equilibrium physics of quantum many-body systems. The theory originates as an analogy to the equilibrium statistical mechanics about the phase transitions \cite{Heyl2013,Heyl2018,Zvyagin2016}. DQPTs can be triggered through a sudden quench protocol from the initially prepared Hamiltonian $H_i$ in one phase or at phase boundary to the final Hamiltonian $H_f$ in another phase. DQPTs occur at times when the dynamic quantity Loschmidt amplitude (LA) defined as
\begin{equation} \label{LA}
	\mathcal{G}(t) = \braket{\psi_0^i | \psi(t)} = \braket{\psi_0^i | e^{-iH_ft / \hbar} | \psi_0^i},
\end{equation}
the overlap of the time-evolved state $\ket{\psi(t)}$ onto the initial ground state $\ket{\psi_0^i}$, vanishes, or equivalently the dynamic free energy or Loschmidt rate (LR)
\begin{equation} \label{LR}
	\lambda(t) = -\lim_{N \rightarrow \infty} \frac{1}{N}\ln[\mathcal{L}(t)],
\end{equation}
where $\mathcal{L}(t) = |\mathcal{G}(t)|^2$ is termed Loschmidt echo (LE), behaves nonanalytically \cite{Heyl2013,Heyl2018,Zvyagin2016}. This is the definition of type-II DQPTs. On the other hand, the type-I DQPTs is defined under the long-term dynamics of a system's order parameter on a sudden quench \cite{Zunkovic2018,Halimeh2017,Yuzbashyan2006,Sciolla2010,Zunkovic2016}. Supported by the realization of DQPTs in experiments for various quantum systems via quantum simulators \cite{Tian2019,Jurcevic2017,Zhang2017,Flaschner2018,Guo2019,Yang2019,Tian2020,Xu2020,Bernien2017,Sanchez2018}, the establishment of the whole DQPT theories thrive. There has been research on the fundamental properties of DQPTs and the theoretical studies of DQPTs in various quantum models \cite{Heyl2013,Heyl2018,Zvyagin2016,Zunkovic2018,Halimeh2017,Yuzbashyan2006,Sciolla2010,Zunkovic2016,Heyl2015,Heyl2014,Weidinger2017,Titum2019,Jafari2019,Schmitt2015,Schmitt2018,Torlai2014,Canovi2014,Sedlmayr2018,Poyhonen2021,Nicola2020,Nicola2021,Vosk2014,HeylPollmann2018,Budich2016,Yu2021,Zache2019,Vajna2014,Lacki2019,Lahiri2019,YangZhou2019,Kosior2018,Kyaw2020,Zvyagin2017,JafariJohannesson2019,Jafari2021,Patra2011,Zamani2020,Bandyopadhyay2021,Halimeh2021,Markov2021,Homrighausen2017,Ding2020,Karrasch2013,Kriel2014,Cheraghi2020,Kennes2018,Seetharam2021,Sadrzadeh2021,Naji2022,Jafari2022,Mishra2020,Halimeh2020,Hashizume2022,Damme2022,Uhrich2020,Lang2018,Andraschko2014,Morrison2008,Porta2020,Wong2022,Corps2022,Corps2023}. DQPTs are also substantially shown to have deep connections to entanglement spectrum and correlation matrix \cite{Heyl2018,Torlai2014,Canovi2014,Sedlmayr2018,Poyhonen2021,Vosk2014,Patra2011,Jafari2021,Morrison2008}.

The wealth of linkage to the entanglement motivates the investigation of a more experimentally accessible quantity regarding spin dynamics, the \textit{spin-squeezing parameter} (SSP). SSP quantifies the quantum fluctuation when one measures one of two non-commutative spin observables and it has various definitions fitting the purpose of studies \cite{Ma2011,Gross2012,Pezze2018}. The idea of using SSP as a tool to study quantum phase transitions was introduced in the past twenty years \cite{Hamley2012,Sun2011,Vidal2004}. A recent paper showed that a certain definition of SSP can be a probe to detect the type-I DQPT, where the SSP behaves nonanalytically against the tunable parameter of a Bose-Einstein condensate after a sudden quench across certain phase boundary \cite{Huang2022}. Similar effect is also observed in experiment for another model between different dynamical phases \cite{Xu2020}. A recent work on SSP in the one-dimensional (1D) XY model, where the authors showed the possibility to generate spin-squeezed state from originally unsqueezed system by quenching, also stated the nonanalyticities found in long-time-averaged SSP against post-quenched Hamiltonian for an arbitrary pre-quenched Hamiltonian, and suggested a potential linkage to equilibrium phase transitions \cite{Cheraghi2022}. These works seem to suggest potential connection between type-I DQPTs and quantum fluctuations, while the short-term physical nature behind the type-II DQPTs is not well-addressed.

Inspired by the extensive use of spin squeezing in the mentioned research, we investigate the spin dynamics by observing the spin squeezing of a spin-$\frac{1}{2}$ model around DQPTs under different quench protocols in this work. We adopt the definition of SSP by Ueda and Kitagawa \cite{Kitagawa1993}
\begin{equation} \label{SSP}
	\xi_S^2 = \frac{4(\Delta J_{\hat{n}_\perp})^2}{N},
\end{equation}
where $(\Delta J_{\hat{n}_\perp})^2$ denotes the variance of the spin operators orthogonal to the mean-spin direction (MSD) of the system,  with the spin operators defined as
\begin{equation} \label{Ja}
	J_{a} = \frac{1}{2} \sum_{j = 1}^N \sigma_j^{a} \qquad a = x,y,z.
\end{equation}
The state is considered to be squeezed if $\xi_S^2<1$, and the larger the SSP is, the less the state is squeezed.

In this work, we combine the conventional LA analysis under a sudden quench and the study of time evolution of SSP defined by Eq. (\ref{SSP}) in the XY model. %\textcolor{red}{We further extend the observation made from the experimental study of type-II DQPTs in transverse-field Ising model \cite{Jurcevic2017}, where the rate of change of SSP attains a maximum at the critical time, to quenches involving more phases and quench types.} 
Along with extracting explicitly the two-point correlation functions in the analytical expression of SSP, we demonstrate how the evolution of quantum fluctuation unveils the spin correlations during DQPT. Namely, our results show the SSP attains local maximum near DQPT in most of the cases. %, which agree with the work in Ref. \cite{Heyl2018}.} 
Furthermore, we show the cause of this increase aligns with the physical properties of the underlying equilibrium phase the system is quenched to. We also distinguish the evolution pattern of SSP for different types of quench, namely quenches across and from the equilibrium phase boundary, and within one phase, in the regime around the first critical time. We believe our study provides insights on filling the knowledge gap of spin fluctuation in DQPTs as well as the physical essence of DQPTs.

The manuscript is organized as follows: Sec. \ref{sec:LRSSP} covers the analytical details necessary for the study, including the expression of LA and SSP. Sec. \ref{sec:results} demonstrates the findings in different quench scenarios, with subsections divided as follows: Sec. \ref{sec:Ising} contains the analysis on quenches between Ising ferromagnetic (FM) and paramagnetic (PM) phases; Sec. \ref{sec:aniso} analyzes the results for quenches between anisotropic phases, and Sec. \ref{sec:multcrit} analysis of quenches from the multicritical point of the XY model. Finally, a conclusion and some future perspectives regarding this work are given in the last section.

\begin{figure} [b!]
	\centering
	\includegraphics[width=8cm]{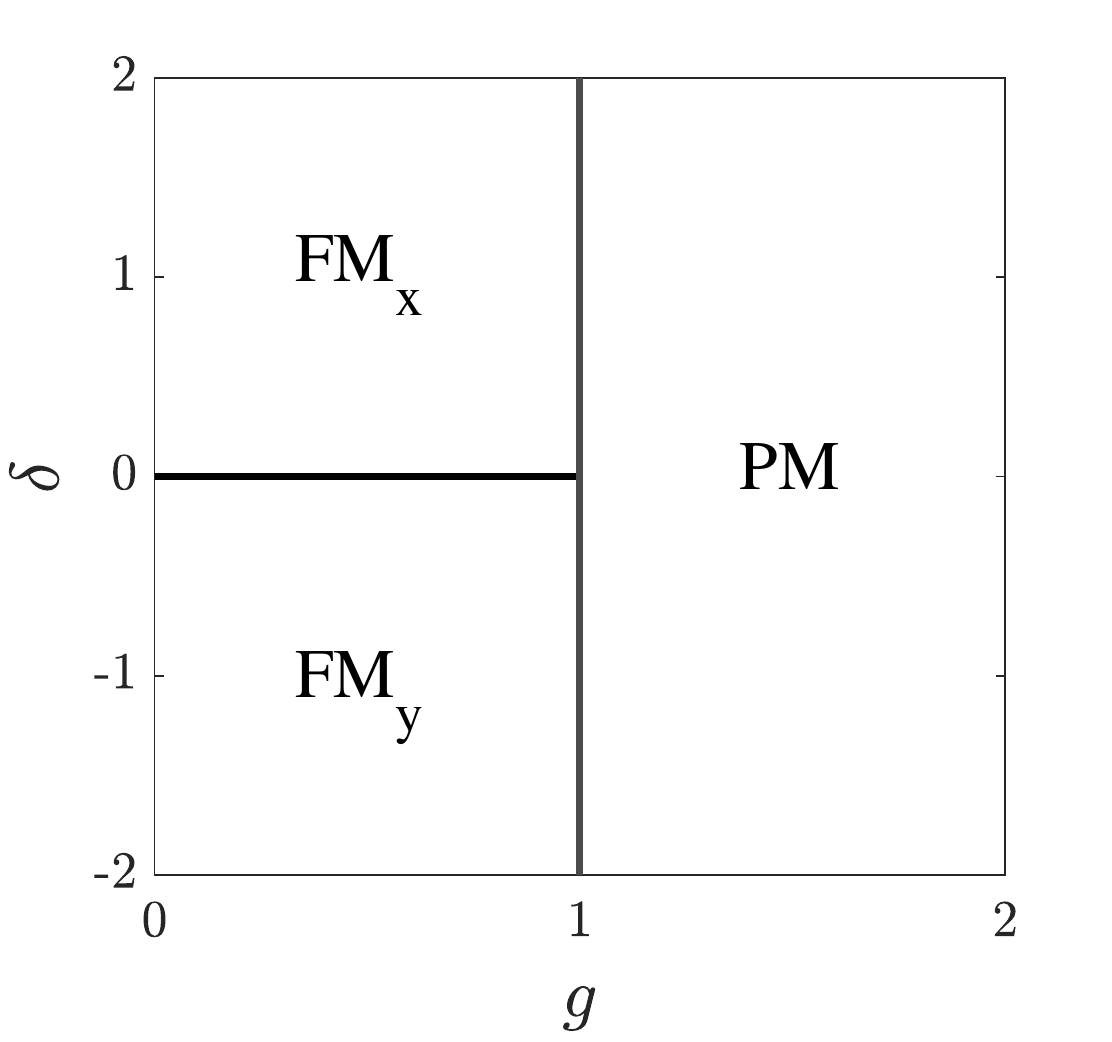}
	\caption{Schematic drawing of the ground-state phase diagram of the equilibrium 1D XY model.}
	\label{XYphase}
\end{figure}

\section{Formulation and 1D XY model}
\label{sec:LRSSP}

We demonstrate our analysis on the 1D XY model which is an integrable model and was proven the existence of DQPTs \cite{Vajna2014,Porta2020}. The Hamiltonian is the following
\begin{equation} \label{XY}
	H(\delta,g) = -\frac{\hbar}{2} \sum_{j = 1}^N \bigg[ J\bigg( \frac{1 + \delta}{2}\sigma_j^x \sigma_{j + 1}^x + \frac{1 - \delta}{2}\sigma_j^y \sigma_{j + 1}^y \bigg) + g\sigma_j^z \bigg],
\end{equation} 
where \textit{N} specifies the system size, \textit{J} is the coupling strength, $\delta$ governs the anisotropic coupling between spins along the \textit{x} and \textit{y} directions, \textit{g} is the external magnetic field strength along the \textit{z} direction, and periodic boundary condition is adopted. Below we set $\hbar = J = 1$ for convenience. The model exhibits competitions between anisotropic and magnetic coupling which results in the existence of multiple phases. In particular, $g < 1$ gives rise to the FM phase, where it is further divided into \textit{x}-spin dominant FM$_x$ phase and \textit{y}-spin dominant FM$_y$ phase depending on the sign of $\delta$. When $g > 1$, the model transits to the PM phase with the ground-state MSD aligning parallel to the external magnetic field. Figure  \ref{XYphase} shows the equilibrium phase diagram of the 1D XY model. We will perform quenches across and from different phase boundaries and critical points.

We study the quench dynamics of the model by first Jordan-Wigner transformation $\sigma_j^+ = \exp[i\pi \sum_{l < j} c_l^\dagger c_l] c_j$ and $\sigma^z_j=1-2c_j^{\dagger}c_j$, followed by a Fourier transformation $c_j = (1/\sqrt{N})\sum_k e^{-ijk} c_k$ \cite{Sachdev,Barouch1970,Barouch1971} and turns the Hamiltonian into
\begin{equation} \label{XYk}
	H(\delta,g) = \sum_{k > 0} \eta_k^\dagger H(k) \eta_k,
\end{equation}
where $\eta_k = (c_k,c_{-k}^\dagger)^T$ and $H(k) = \vec{d}(k) \cdot \vec{\sigma}$ is the Bloch Hamiltonian with the Bloch vector $\vec{d}(k) = (0,-\delta\sin k,g - \cos k)^T$ and $\vec{\sigma}=(\sigma^x,\sigma^y,\sigma^z)$ are the Pauli matrices \cite{Budich2016}. For the sake of analysis, we concern the unit Bloch vectors $\hat{d}(k) = \vec{d}(k) / |\vec{d}(k)|$. We restrict our calculations in the even-parity subspace and $N$ being even with the \textit{k}'s of the form $k = (2m-1)\pi / N$, where $m = 1,2,\dots,N/2$. Performing the Bogoliubov transformation $c_k = \cos[\theta_k(\delta,g)] \beta_k + i\sin[\theta_k(\delta,g)] \beta_{-k}^\dagger$, the diagonalized Hamiltonian reads
\begin{equation} \label{diagH}
	H(\delta,g) = \sum_{k > 0} \varepsilon_k(\delta,g)( \beta_k^\dagger \beta_k - \beta_{-k} \beta_{-k}^\dagger ),
\end{equation}
where $\varepsilon_k(\delta,g) = |\vec{d}(k)| = \sqrt{( \cos k - g )^2 + ( \delta\sin k )^2}$ is the quasiparticle eigenenergy, and $\theta_k(\delta,g) \in [0,\pi/2]$ is the Bogoliubov angle defined such that $\tan[ 2\theta_k(\delta,g) ] \equiv \delta\sin k / ( \cos k - g )$.

Consider the system initially prepared in $H_i = H(\delta_i,g_i)$ and quenched to $H_f = H(\delta_f,g_f)$. Using the vacuum state as the initial state, LA in Eq.(\ref{LA}) can be decomposed into momentum-wise form, namely $\mathcal{G}(t) = \prod_{k > 0} \mathcal{G}_k(t)$, where
\begin{equation} \label{Gk}
	\mathcal{G}_k(t) = \cos^2\phi_k e^{i\varepsilon_k(\delta_f,g_f)t} + \sin^2\phi_k e^{-i\varepsilon_k(\delta_f,g_f)t}
\end{equation}
with $\phi_k = \theta_k^i - \theta_k^f$ being the difference between the initial and the final Bogoliubov angles. The nonanalyticities in LA occurs when $\phi_k = \pm\pi/4$ at critical times
\begin{equation} \label{ct}
	t_c^p = \frac{\pi}{\varepsilon_{k^*}^f}\bigg( p - \frac{1}{2} \bigg) \qquad p \in \mathbb{N},
\end{equation}
where $k^*$ is the \textit{critical momentum} when the corresponding initial and final unit Bloch vectors $\hat{d}_i(k^*)$ and $\hat{d}_f(k^*)$ are orthogonal to each other, i.e. $\hat{d}_i(k^*) \cdot \hat{d}_f(k^*) = 0$.

The behavior of the momentum-wise LA can be visualized by the vector introduced by Ding \cite{Ding2020}:
%The dynamics in Bloch space can be visualized by the vector defined by Ding \cite{Ding2020}
\begin{equation} \label{rvector}
	\vec{r}_k(t) = (x_k,y_k) = |\mathcal{G}_k(t)| e^{i\phi_k^G(t)},
\end{equation}
where $\phi_k^G(t) = \phi_k^\mathcal{G}(t) - \phi_k^\text{dyn}(t)$ is the Pancharactnam geometric phase (PGP) \cite{Pancharatnam1956,Samuel1988}, defined by the difference between the phase of LA $\phi_k^\mathcal{G}(t) = \text{Arg}[\mathcal{G}_k(t)]$ and the dynamical phase $\phi_k^\text{dyn}(t)=-\int_0^t dt' \braket{\psi_k(t') | H_f(k) | \psi_k(t')}=\varepsilon_k^f \cos( 2\phi_k ) t $. The PGP characterises the dynamical topological order parameter (DTOP) \cite{Budich2016}
\begin{equation} \label{DTOP}
	\nu_D = \frac{1}{2\pi} \oint_0^\pi \frac{\partial \phi_k^G(t)}{\partial k},
\end{equation}
which changes its value at DQPTs.

%which is defined by the difference between the phase of LA $\phi_k^\mathcal{G}(t) = \text{Arg}[\mathcal{G}_k(t)]$ and the dynamical phase
%\begin{equation} \label{phikdyn}
%	\begin{aligned}
%		\phi_k^\text{dyn}(t) & = -\int_0^t dt' \braket{\psi_k(t') | H_f(k) | \psi_k(t')} \\
%		& = \varepsilon_k^f \cos( 2\phi_k ) t.
%	\end{aligned}
%\end{equation}
%This quantity captures how the momentum-wise LA behaves far from equilibrium, especially around the critical momentum $k^*$.

On the other hand, we adopt the formulation in Ref. \cite{Cheraghi2022} regarding the spin dynamics analysis. The MSD of XY model is along \textit{z} direction as a result of the $\mathbb{Z}_2$ symmetry which guarantees $\braket{J_x} = \braket{J_y} = 0$. The factor $(\Delta J_{\hat{n}_\perp})^2$ in Eq. (\ref{SSP}) is then taken to be $[(\cos\alpha) J_x + (\sin\alpha) J_y]^2$. Setting $\braket{\mathcal{O}} \equiv \braket{\psi(t) | \mathcal{O} | \psi(t)}$ for any operator $\mathcal{O}$ for simplicity, the SSP reads
\begin{equation} \label{TSSP}
	\xi_S^2(t) = \frac{2}{N}\min\{ \braket{J_x^2 + J_y^2} + J(\alpha,t) \},
\end{equation}
where
\begin{equation} \label{Jalpha}
	J(\alpha,t) = \cos( 2\alpha ) \braket{J_x^2 - J_y^2} + \sin( 2\alpha ) \braket{J_x J_y + J_y J_x}.
\end{equation}
The angle which yields the minimum of Eq. (\ref{TSSP}) is termed the \textit{squeezing angle} and is denoted as $\alpha_s$ in the following. This quantity has been studied in the context of qubit systems relating its time evolution to the energy spectrum of the systems \cite{Jin2007}. In this paper, we will demonstrate how it embeds the competition between the net parallel-spin contribution and the cross-spin contribution over time. With transitional symmetry of the chain, the time evolution of SSP can be written as
\begin{equation} \label{SSPA}
	\begin{aligned}
		\xi_S^2(t) & = 1 + 2\sum_{n = 1}^{N - 1} ( G_n^{xx}(t) + G_n^{yy}(t) ) \\
		& \quad - 2\Bigg\{\bigg[ \sum_{n = 1}^{N - 1} ( G_n^{xx}(t) - G_n^{yy}(t) ) \bigg]^2 \\
		& \quad + \bigg[ \sum_{n = 1}^{N - 1} ( G_n^{xy}(t) + G_n^{yx}(t) ) \bigg]^2\Bigg\}^{\frac{1}{2}},
	\end{aligned}
\end{equation}
where $G_n^{ab}(t) = \braket{\sigma_1^a \sigma_{1 + n}^b}/4$ is the correlation function along direction \textit{a} and \textit{b} and $G_0^{aa}(t) = \braket{\sigma_1^a \sigma_1^a}/4 = 1/4$.

The calculation of the correlation functions involves computing strings of operators of the forms \cite{Lieb1961}
\begin{equation} \label{Gab}
	\begin{aligned}
		G_n^{xx}(t) & = \frac{1}{4}\braket{B_1 A_2 B_2 \cdots A_n B_n A_{n + 1}} \\
		G_n^{yy}(t) & = \frac{(-1)^n}{4}\braket{A_1 B_2 A_2 \cdots B_n A_n B_{n + 1}} \\
		G_n^{xy}(t) & = \frac{i}{4}\braket{B_1 A_2 B_2 \cdots A_n B_n B_{n + 1}} \\
		G_n^{yx}(t) & = \frac{i(-1)^{n - 1}}{4}\braket{A_1 B_2 A_2 \cdots B_n A_n A_{n + 1}},
	\end{aligned}
\end{equation}
where $A = c_j^\dagger + c_j$ and $B = c_j^\dagger  - c_j$ in Jordan-Wigner basis. By Wick's theorem, the expectation values turn into sums of all possible permutations of products of operator pairs, which turns out to be a problem of calculating pfaffians of $2n \times 2n$ skew-symmetric matrices
\begin{equation} \label{pf}
	G_n^{ab}(t) \sim \text{pf}
	\begin{pmatrix}
		\braket{\mathcal{O}_1 \mathcal{O}_2} & \braket{\mathcal{O}_1 \mathcal{O}_3} & \braket{\mathcal{O}_1 \mathcal{O}_4} & \cdots & \braket{\mathcal{O}_1 \mathcal{O}_{2n}} \\
		& \braket{\mathcal{O}_2 \mathcal{O}_3} & \braket{\mathcal{O}_2 \mathcal{O}_4} & \cdots & \braket{\mathcal{O}_2 \mathcal{O}_{2n}} \\
		& & \braket{\mathcal{O}_3 \mathcal{O}_4} & \cdots & \braket{\mathcal{O}_3 \mathcal{O}_{2n}} \\
		& & & \ddots & \vdots \\
		& & & & \braket{\mathcal{O}_{2n - 1} \mathcal{O}_{2n}}
	\end{pmatrix}.
\end{equation}
Note that at $t = 0$ for parallel-spin correlation $a = b = x,y$, the pfaffians reduce to Toeplitz determinants known for the analytical works of Lieb \textit{et al.} in correlation function calculations in XY model \cite{Lieb1961}. The elements of the pfaffians are four types of two-point functions as follows:
\begin{equation} \label{2ptfunc}
	\begin{aligned}
		\braket{A_j A_{j + n}} & = \braket{B_j B_{j + n}} \\
		&= -i\frac{2}{N} \sum_{k > 0} \sin( 2\phi_k ) \sin( nk ) \sin( 2\varepsilon_k^ft ) \\
		\braket{A_j B_{j + n}} & = \frac{2}{N} \sum_{k > 0} [ \cos( 2\phi_k ) \cos( 2\theta_k^f + nk ) \\
		& \quad - \sin( 2\phi_k ) \sin( 2\theta_k^f + nk ) \cos( 2\varepsilon_k^ft ) ] \\
		\braket{B_j A_{j + n}} & = -\frac{2}{N} \sum_{k > 0} [ \cos( 2\phi_k ) \cos( 2\theta_k^f - nk ) \\
		& \quad - \sin( 2\phi_k ) \sin( 2\theta_k^f - nk ) \cos( 2\varepsilon_k^ft ) ].
	\end{aligned}
\end{equation}

In the next section, we study the spin dynamics in terms of the SSP and the correlation functions in relation to DQPT under three scenarios:
\begin{enumerate}
	\item quenches between Ising phases, where we fixed the $\delta$ and alter along \textit{g};
	\item quenches between anisotropic phases, where we fixed \textit{g} and quench along $\delta$ in the region bounded by $-\infty < \delta < \infty$ and $0 \leq g < 1$;
	\item quenches from the multicritical point $(\delta_c,g_c) = (0,1)$.
\end{enumerate}
Alternatively we use the shorthanded notation $(\delta_i,g_i) \rightarrow (\delta_f,g_f)$ for any quench when necessary. All quenches were performed on an $N = 100$ system, and we concentrate our discussions around the first critical time.

\section{Dynamics of spin squeezing}
\label{sec:results}

%In this section, we demonstrate all the exact results of the quenches between three different phases: FM$_x$ phase, FM$_y$ phase and PM phase.

\subsection{Quenches between Ising phases}
\label{sec:Ising}

\begin{figure} [t!]
	\centering
	\includegraphics[width=8.5cm]{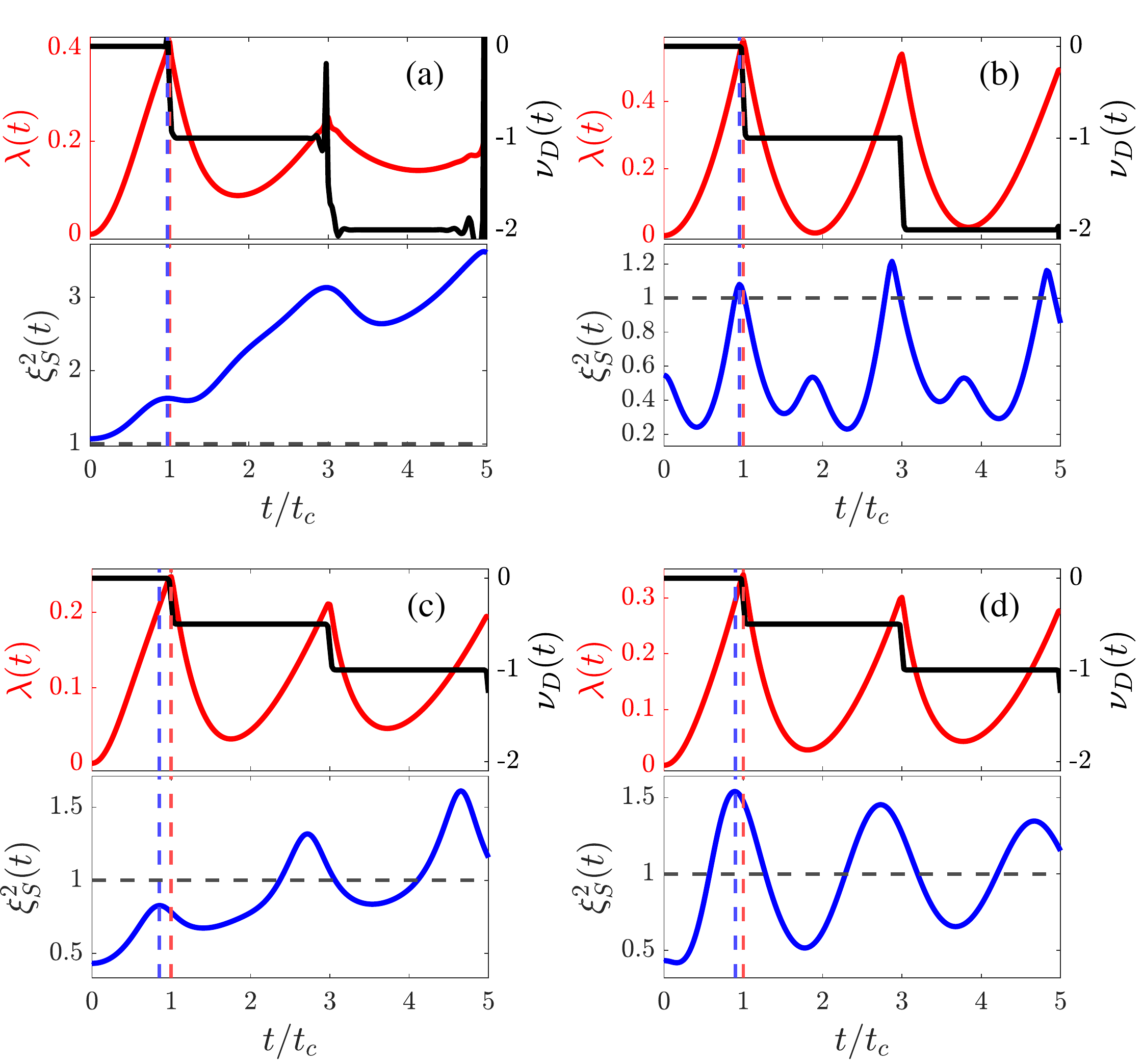}
	\caption{Scaled time plots of LR, DTOP (top panels) and SSP (bottom panels) at $\delta = 0.8$ for (a) $g_i = 0.4 \rightarrow g_f = 2$, (b) $g_i = 2 \rightarrow g_f = 0.1$, (c) $g_i = 1 \rightarrow g_f = 2$ and (d) $g_i = 1 \rightarrow g_f = 0.5$. Red and blue dashed lines indicate the first peak of LRs and SSPs respectively. Black dashed horizontal lines indicate $\xi_S^2(t) = 1$.}
	\label{IsingLRSSP}
\end{figure}

\begin{figure} [t!]
	\centering
	\includegraphics[width=8.5cm]{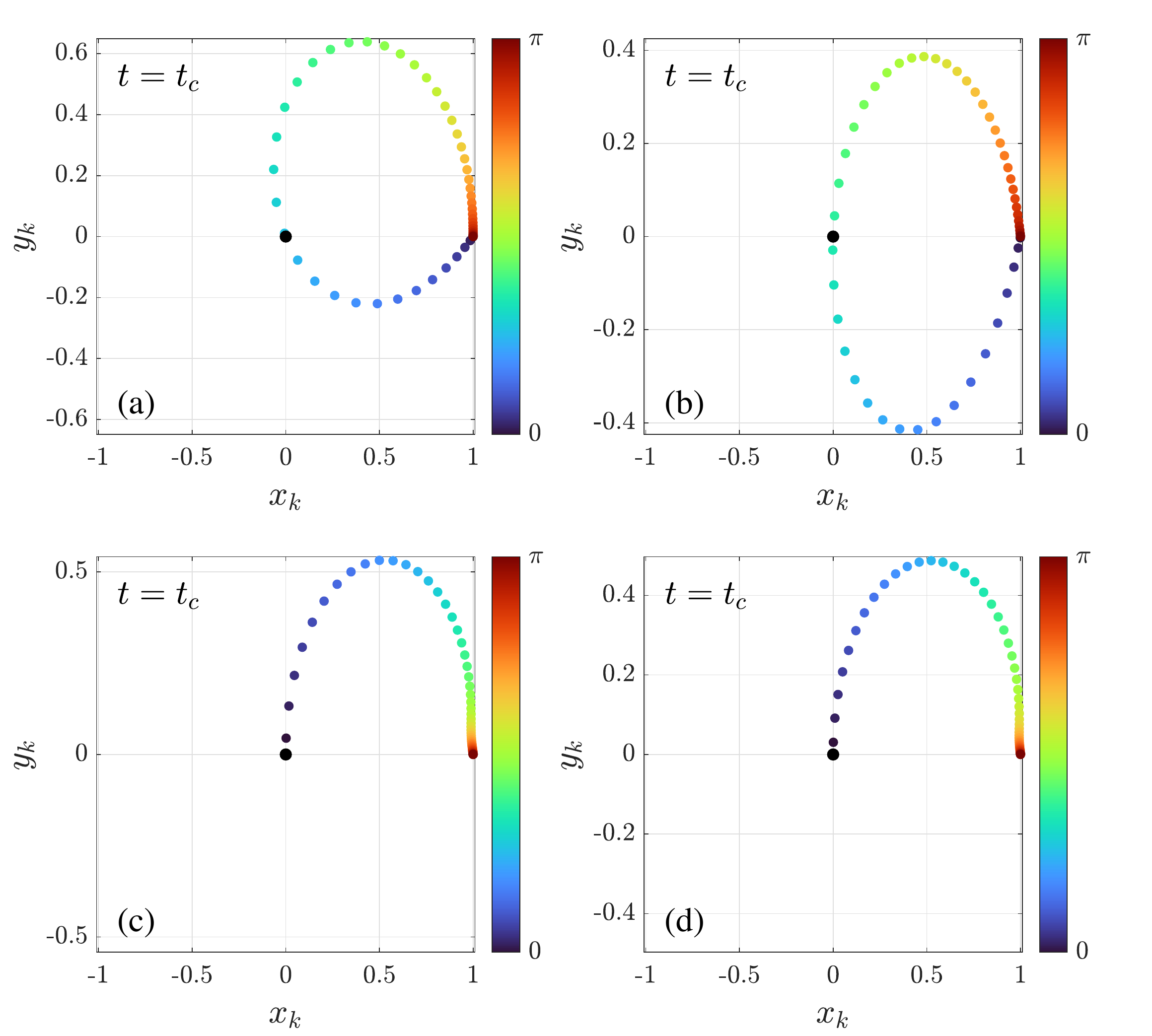}
	\caption{Trajectories of $\vec{r}_k$'s at the corresponding critical times for the same quench cases in FIG. \ref{IsingLRSSP}. Black dots indicate the origin the vectors are located. The color scale shows the progression of $\vec{r}_k$'s from $k = 0$ to $\pi$.}
	\label{Isingrk}
\end{figure}

\begin{figure} [t!]
	\centering
	\includegraphics[width=8.5cm]{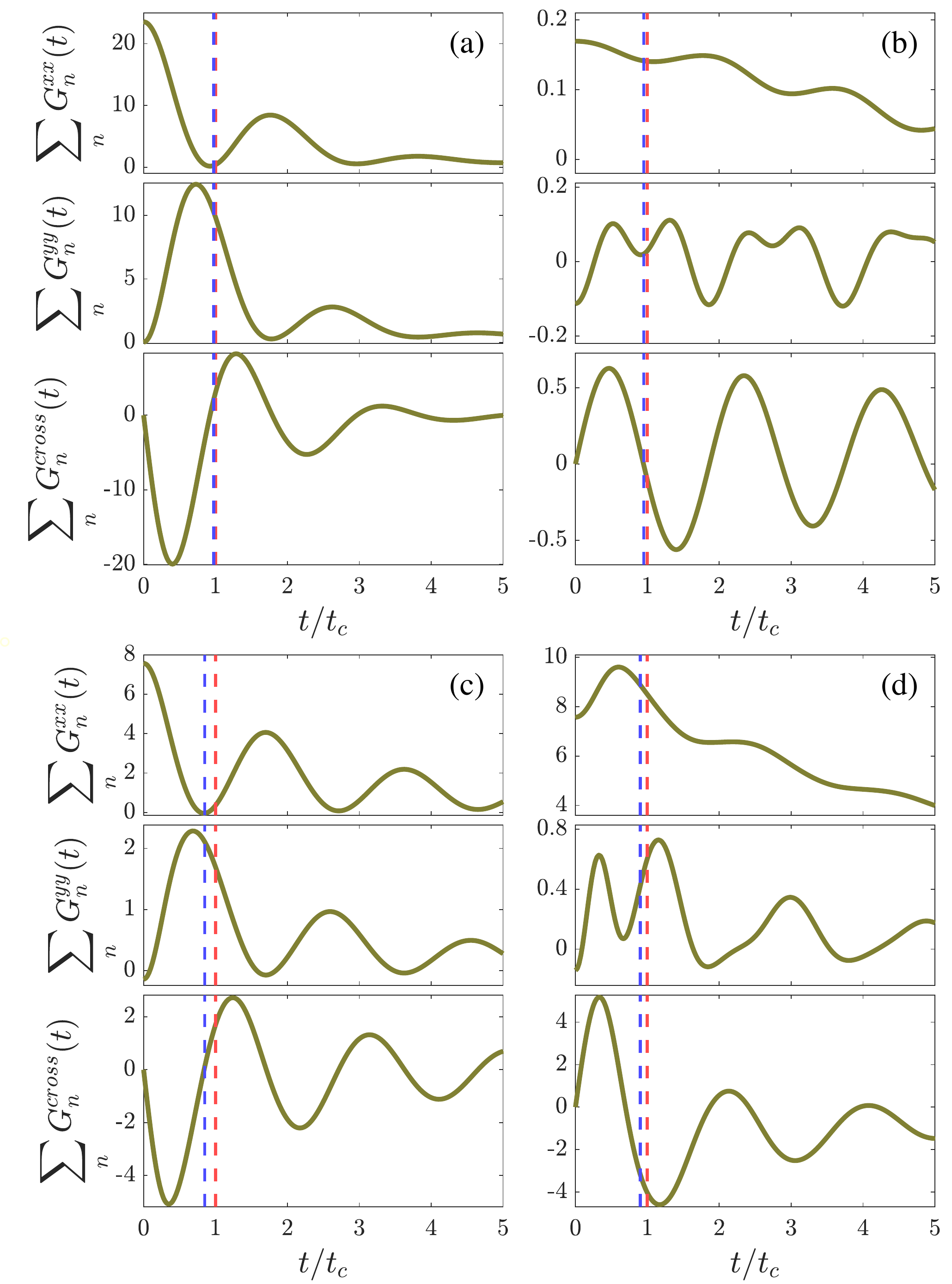}
	\caption{Scaled time plots of sums of correlation functions $G_n^{xx}(t)$ (top panels), $G_n^{yy}(t)$ (middle panels) and $G_n^{cross}(t)$ (bottom panels) for the same quench cases in FIG. \ref{IsingLRSSP}. Red and blue dashed lines indicate the first peak of LRs and SSPs respectively.}
	\label{IsingCorr}
\end{figure}

\begin{figure} [t!]
	\centering
	\includegraphics[width=8.5cm]{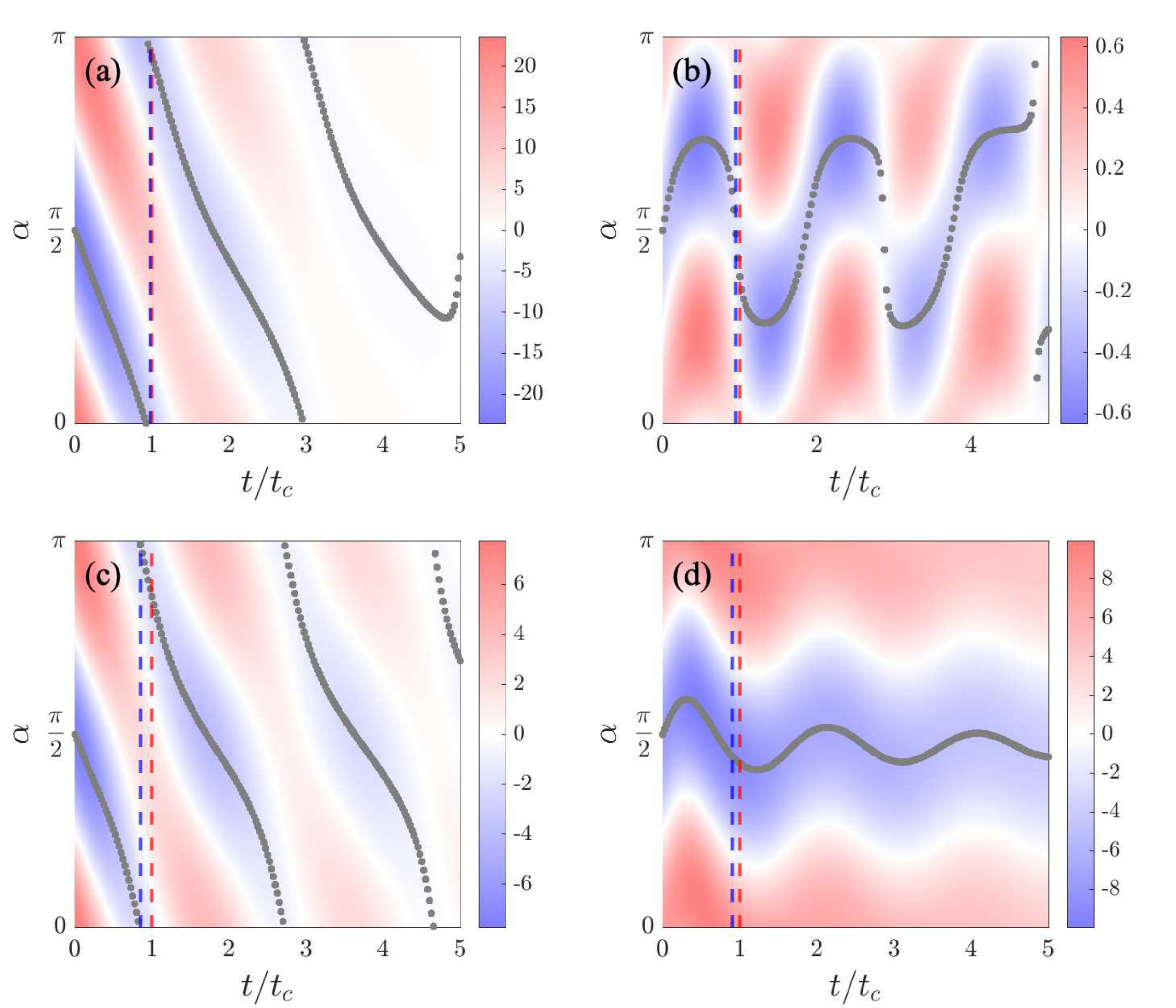}
	\caption{Colormap of $J(\alpha,t)$ for the same quench in FIG. \ref{IsingLRSSP}. Grey dots emphasize the squeezing angle $\alpha_s$. Red and blue dashed lines refers to the first peak of LRs and SSPs respectively.}
	\label{Isingalpha}
\end{figure}

We performed quench cases involving the Ising boundary $g_c = 1$ while fixing $\delta = 0.8$. The quench dynamics of forward quench ($g_i < g_f$), backward quench ($g_i > g_f$) and quench from critical point ($g_i = g_c = 1$) are shown in FIG. \ref{IsingLRSSP}. The DTOPs show the discontinuous jumps at critical times as expected. Note that all the extra spikes on the DTOPs in this paper are finite-size effect as confirmed numerically and they would vanish to give true DTOP evolution pattern shown in Ref. \cite{Budich2016} for larger systems. The vectors $\vec{r}_k$'s plotted in FIG. \ref{Isingrk}a and \ref{Isingrk}b swipe a full circle passing through the origin for quenches across the equilibrium phase boundary. The momentum at which the vector is zero is the critical momentum, where $\mathcal{G}_{k^*}(t_c) = 0$. $\vec{r}_k$'s only swipe half a circle for quenches from the critical point, which is consistent with the observation in Ref. \cite{Ding2020} and this leads to the half-integer jump in DTOP.

From FIG. \ref{IsingLRSSP}, we observe that the peaks of SSPs in whichever case occurs slightly sooner than the DQPTs. Note that the critical times indicated in the figures are calculated by Eq. (\ref{ct}) in the thermodynamic limit. For system sizes $N \ge 100$, the peak times extracted from SSP do not change significantly, suggesting this time discrepancy remains in larger systems. Nevertheless, the SSP is a local maximum, which implies the state is the least spin-squeezed, in the vicinity of DQPT. The difference between forward and backward quenches is the SSP keeps increasing for forward quenches (left column of FIG. \ref{IsingLRSSP}) while it stays near and below 1 for backward quenches (right column of FIG. \ref{IsingLRSSP}) , but they both have some revival effects during DQPT where the spin-squeezing parameter falls following its peak, decreasing in parallel to the Loschmidt rate. This suggests that  the system is around its most uncertain spin state along the direction perpendicular to the MSD around the critical time.

The evolution pattern of SSPs can be qualitatively understood via the summed time-evolved correlation functions and squeezing angle, which are plotted in FIG. \ref{IsingCorr} and \ref{Isingalpha}, respectively. For forward quenches including quenches from Ising boundary (FIG. \ref{IsingCorr}a and \ref{IsingCorr}c), the $x$-direction correlation dominates initially as expected. As one approaches the first critical time, the \textit{y}-direction correlation functions surpass the \textit{x}-direction correlations. At the same time, the effect of cross-direction correlations $G_n^{cross}(t) = G_n^{xy}(t) + G_n^{yx}(t)$ grows rapidly. In the vicinity of the first DQPT, this surpassing effect is among the highest while the contribution from the cross-direction correlations sinks to near zero. After the DQPT, $G_n^{xx}(t)$'s revive partly, and both $G_n^{xx}(t)$'s and $G_n^{yy}(t)$'s have small but non-zero contributions while the cross-spin correlations oscillate around zero in the later time. This phenomenon resonates with the fact that the quench is from FM phase, which is \textit{xx}-correlation dominant, to PM phase, in which none of \textit{x}- and \textit{y}-correlations dominate. Around DQPT, the system not only stays temporarily least spin-squeezed, but also the furthest away from the original phase where \textit{yy}-correlations thrive and surpass \textit{xx}-correlations.

In terms of the squeezing angle, we see from Eq. \ref{Jalpha} that $J(\alpha_s,t)$ exhibits dynamics resembling the rotation of a vector during quenching. In particular, we define a \textit{squeezing vector} with \textit{x} component $\braket{J_x^2 - J_y^2}$ and \textit{y} component $\braket{J_x J_y + J_y J_x}$ and an angle $2\alpha_s$ between them. By observing the evolution of the squeezing angle $\alpha_s$ as indicated by the grey dots in the plots, the vector has "rotated" in the clockwise direction for about half a circle before reaching the first critical time. The change of $\alpha_s$ seems to be the most rapid across DQPT. After one revival of the LE where the LR attains a local minimum (at $t/t_c\approx 2$), the vector has completed about a full revolution.

%the vector ``rotates" a full revolution in clockwise direction after one revival of the Loschmidt rate, \textcolor{red}{whilst at DQPT the change of $\alpha$ seems to be the most rapid, as shown in FIG. \ref{Isingalpha}a and \ref{Isingalpha}c. The dynamical phase transition occurs around the point that the squeezing vector rotates half a circle.}

On the other hand, from FIG. \ref{IsingCorr}b and \ref{IsingCorr}d, backward quenches seem to align with our observation to the spin nature of the system established before: The correlation functions along \textit{x} direction are the strongest alongside with a local minimum of \textit{y}-direction correlations in the vicinity of DQPT, marking the temporary transition to a state with FM$_x$ character. Apart from that, the contribution from the cross-spin correlations briefly vanishes. In fact, in all quenches except those from Ising boundary the $G_n^{cross}(t)$'s change their sign when passing the critical time. The sign change is opposite to that for forward quenches. Another qualitative difference from forward quenches is the rotation direction of the squeezing vector, where it rotates anticlockwise initially and the motion oscillate around $\alpha = \pi/2$ (see FIG. \ref{Isingalpha}b and \ref{Isingalpha}d). Like the forward quenches, the rate of change of $\alpha_s$ around DQPT is the strongest and the features are more apparent for ``longer" ($\Delta g = g_f - g_i$ is large) quench.

\subsection{Quenches between anisotropic phases}
\label{sec:aniso}

\begin{figure} [t!]
	\centering
	\includegraphics[width=8.5cm]{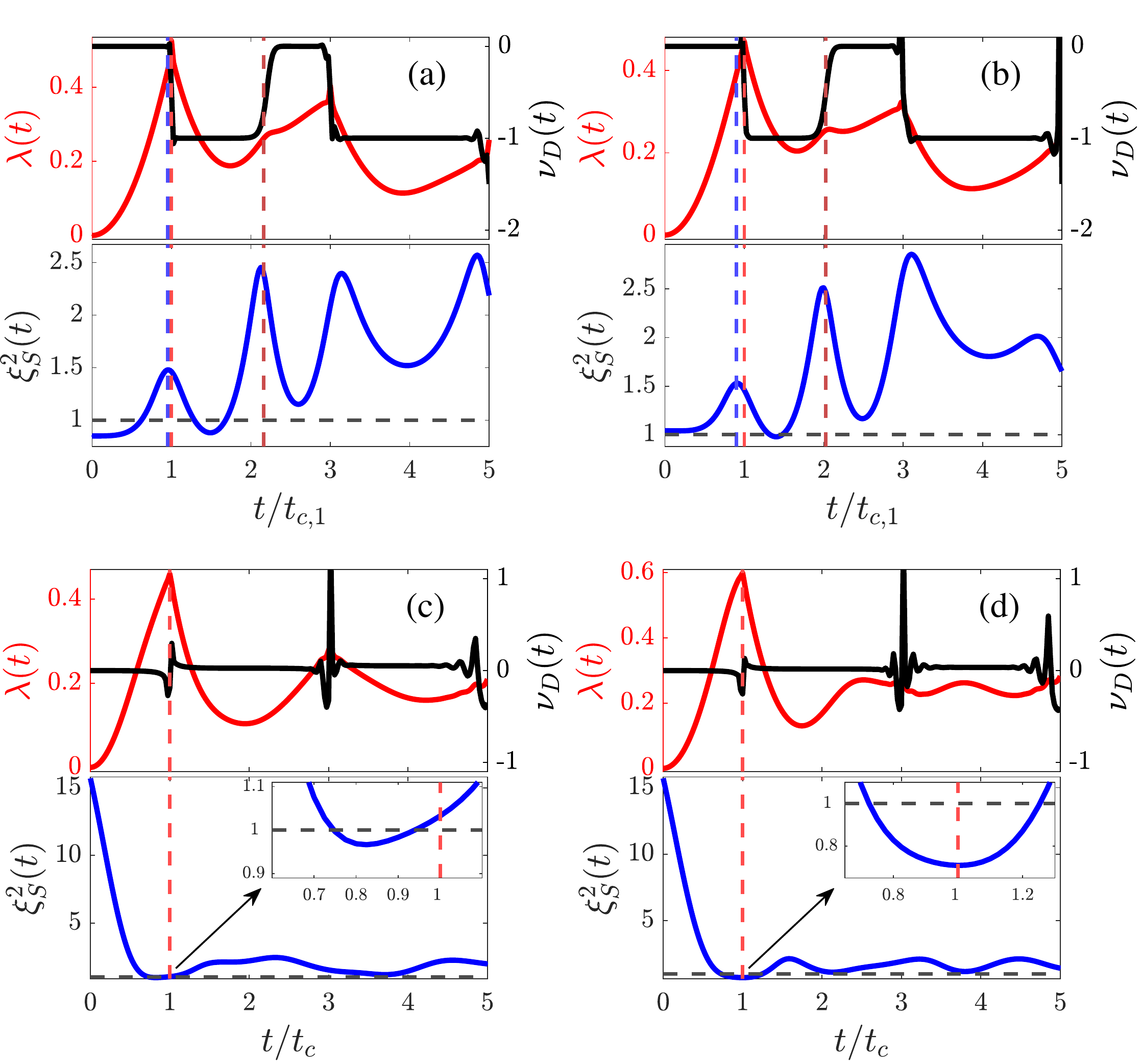}
	\caption{Scaled time plots of LR, DTOP (top panels) and SSP (bottom panels) at $g = 0.5$ for (a) $\delta_i = -1.2 \rightarrow \delta_f = 0.8$, (b) $\delta_i = 0.8 \rightarrow \delta_f = -0.8$, (c) $\delta_i = 0 \rightarrow \delta_f = 0.8$ and (d) $\delta_i = 0 \rightarrow \delta_f = -1.2$. Red and blue dashed lines indicate the first peak of LRs and SSPs respectively. The second red dashed lines in (a) and (b) represents the second critical time $t_{c,2}$ for each quench. Black dashed horizontal lines indicate $\xi_S^2(t) = 1$. The inset plots in (c) and (d) displays the zoomed SSP around the first critical time.}
	\label{AnisoLRSSP}
\end{figure}

\begin{figure} [t!]
	\centering
	\includegraphics[width=8.5cm]{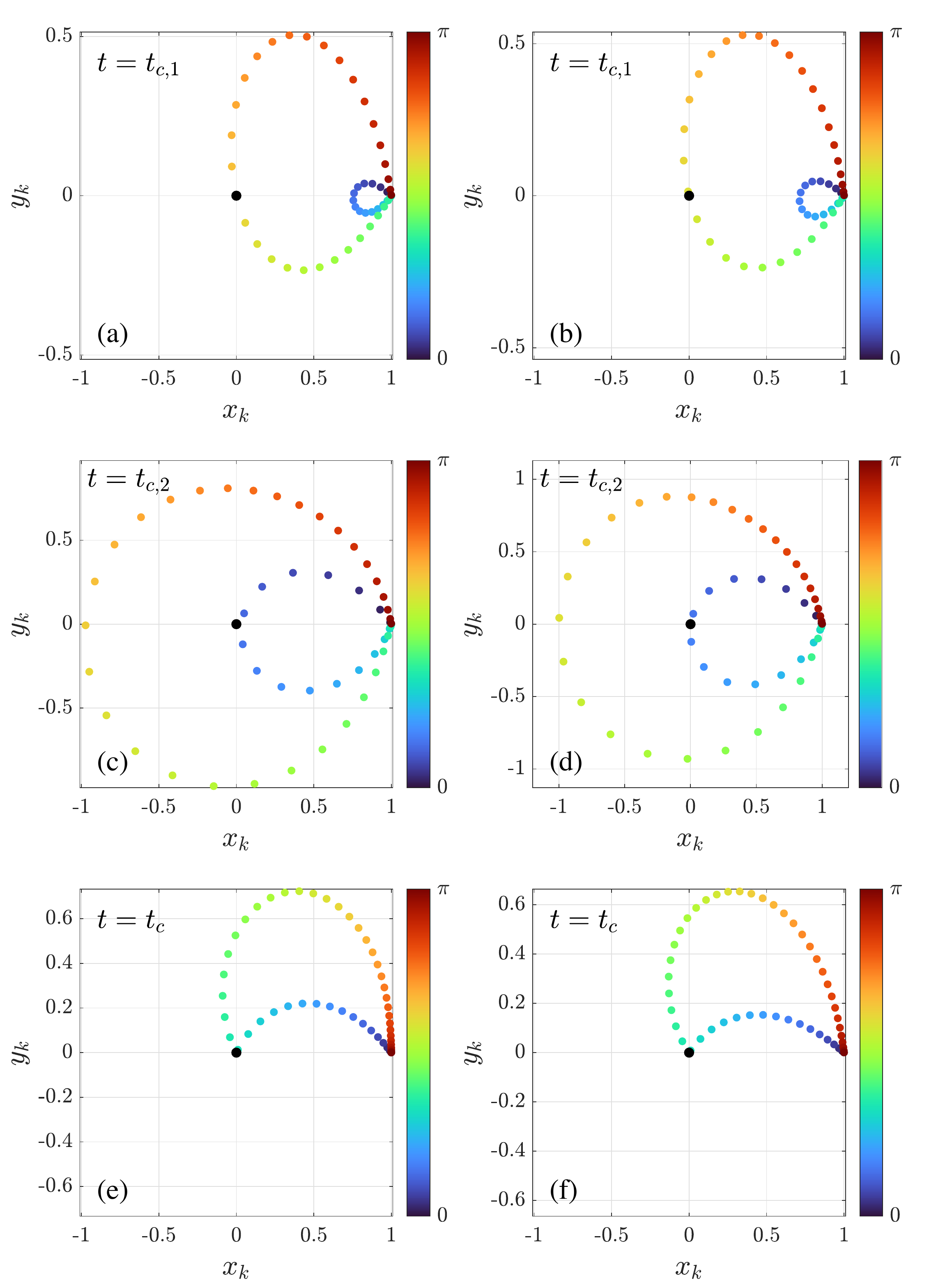}
	\caption{Trajectories of $\vec{r}_k$'s at the critical times: (a,c) upward quench (FIG. \ref{AnisoLRSSP}a), (b,d) downward quench (FIG. \ref{AnisoLRSSP}b), quenches from anisotropic boundary to (e) FM$_x$ (FIG. \ref{AnisoLRSSP}c) and (f) FM$_y$ (FIG. \ref{AnisoLRSSP}d) phase. Black dots indicate the origin the vectors are located. The color scale shows the progression of $\vec{r}_k$'s from $k = 0$ to $\pi$.}
	\label{Anisork}
\end{figure}

\begin{figure} [t!]
	\centering
	\includegraphics[width=8.5cm]{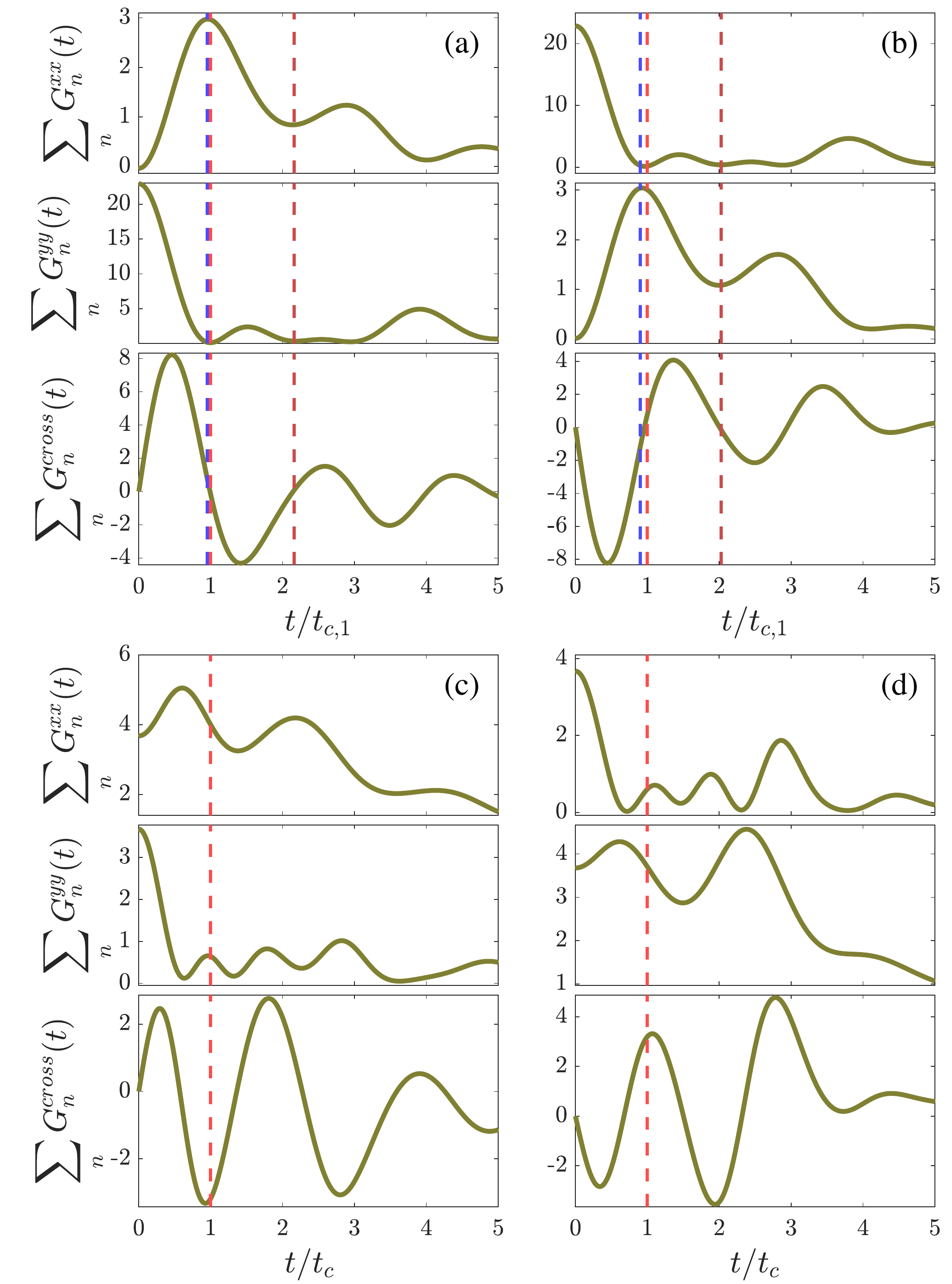}
	\caption{Scaled time plots of sums of correlation functions $G_n^{xx}(t)$ (top panels), $G_n^{yy}(t)$ (middle panels) and $G_n^{cross}(t)$ (bottom panels) for the same quench cases in FIG. \ref{AnisoLRSSP}. Red and blue dashed lines indicate the first peak of LRs and SSPs respectively. The second red dashed lines in (a) and (b) represents the second critical time $t_{c,2}$ for each quench.}
	\label{AnisoCorr}
\end{figure}

\begin{figure} [t!]
	\centering
	\includegraphics[width=8.5cm]{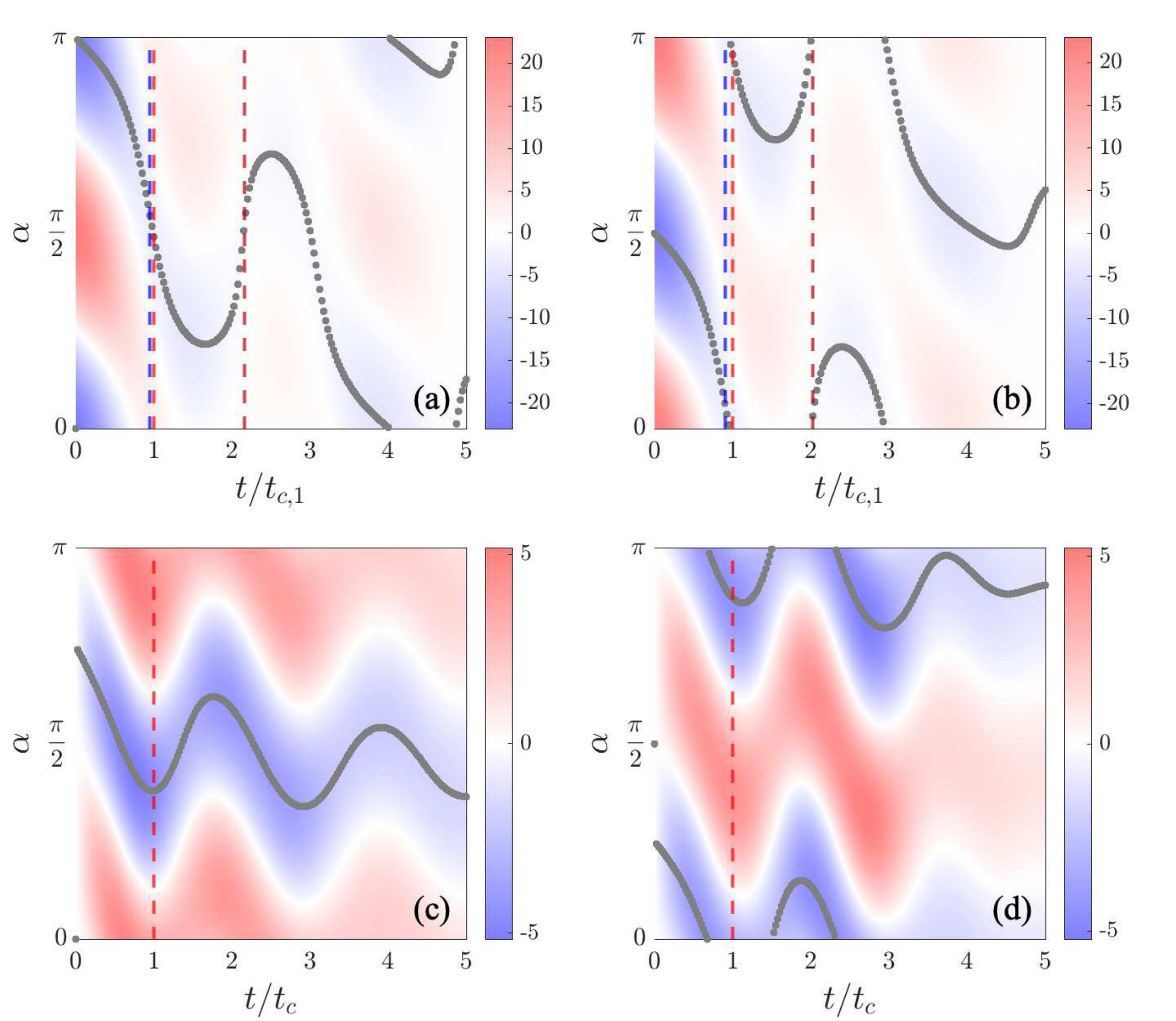}
	\caption{Colormap of $J(\alpha,t)$ for the same quench cases in FIG. \ref{AnisoLRSSP}. Grey dots emphasize the squeezing angle $\alpha_s$. Red and blue dashed lines refers to the first peak of LRs and SSPs respectively. The second red dashed lines in (a) and (b) represents the second critical time $t_{c,2}$ for each quench.}
	\label{Anisoalpha}
\end{figure}

The dynamics of the system in quenches between the anisotropic phases behaves differently from that between Ising phases. One major difference is that there will be two critical momenta $k_1^*$ and $k_2^*$ whenever one quenches across the anisotropic boundary, which in turn give two critical times $t_{c,1}$ and $t_{c,2}$. Budich and Heyl showed that at the two critical momenta, the Bloch vectors rotate in opposite direction when passing through the respective critical time \cite{Budich2016}. Here we demonstrate this idea through the vector defined in Eq. (\ref{rvector}) and further explore the system's dynamics through the spin-squeezing parameter.

The anisotropic boundary is at $\delta_c = 0$, separating two phases -- the \textit{x}-spin dominant FM$_x$ phase ($\delta > 0$) and the \textit{y}-spin dominant FM$_y$ phase ($\delta < 0$). We fix \textit{g} to 0.5 and quench from FM$_y$ to FM$_x$ phase (upward quench), FM$_x$ to FM$_y$ phase (downward quench) and also from the equilibrium phase boundary. Their dynamics are shown in FIG. \ref{AnisoLRSSP}. Clearly from the DTOPs in FIG. \ref{AnisoLRSSP}a and \ref{AnisoLRSSP}b, we can observe the negative and positive jumps at $t_{c,1}$ and $t_{c,2}$ respectively. For each critical time, there is an SSP peak slightly before it, which is the same as those quenches in the case involving the Ising phases in Sec. \ref{sec:Ising}. Notice that the upward and downward quenches are similar for a fixed \textit{g}, unlike the forward and backward quenches in Ising scenarios.

The existence of two critical momenta can be realized in the trajectory of $\vec{r}_k$. Namely, there are two ``loops" in the trajectory plot. %where the dots line from dark blue to bluish green for one loop, then light green to dark red for the second loop. 
At whichever critical time, the corresponding vector $\vec{r}_{k^*}$ vanishes. FIG. \ref{Anisork}a and \ref{Anisork}b demonstrate the phenomena at $t_{c,1}$, the critical time for $k_1^*$. When the system keeps evolving, the blue loop would expand further and it touches the origin at $t_{c,2}$, the critical time for $k_2^*$ (as shown in FIG. \ref{Anisork}c and FIG. \ref{Anisork}d). The opposite circular motion of the trajectories provides a visualization of the integer jumps in DTOP when passing through critical times.

On the other hand, the dynamics of quench from the anisotropic boundary is qualitatively very different from other quenches. The DTOP behaves not as expected (see the top panel of FIG. \ref{AnisoLRSSP}c and \ref{AnisoLRSSP}d), it has nonanalyticities at critical times, yet it does not jump between adjacent DQPTs. This can be seen by the $\vec{r}_k$ in FIG. \ref{Anisork}e and \ref{Anisork}f, where the trajectory does not change smoothly at DQPT when crossing the critical momentum, but the angles do. The evolution of SSP goes from a very high value to a minimum around DQPT. Unlike other quench scenarios, the system is among the most squeezed state near DQPT. The cause of this might be the equilibrium SSP in the domain $\delta = 0,g \in [0,1)$ diverges. When quenched out of the phase boundary, the SSP falls rapidly and reaches its minimum near DQPT.

Since anisotropic phases involves competition between \textit{x}- and \textit{y}-spin interactions, a more direct interpretation about what is happening during DQPT via the time evolution of the summed correlation functions can be observed in FIG. \ref{AnisoCorr}. In particular, the upward quenches in FIG. \ref{AnisoCorr}a do show an increase in $G_n^{xx}(t)$ and a decrease in $G_n^{yy}(t)$, and they reach maximum and minimum respectively in the vicinity of first DQPT. The downward quenches in FIG. \ref{AnisoCorr}b shows the opposite phenomenon. In both cases, the contribution of cross-spin correlations vanish at the first critical times. One observes the same effect for quenches from anisotropic boundary, where the \textit{xx} correlations reach a higher value whilst \textit{yy} correlations drop to near zero (see FIG. \ref{AnisoCorr}c). The feature appearing only in quenches from anisotropic boundary is that at DQPT, the effect of cross-direction correlation functions reach their extrema (FIG. \ref{AnisoCorr}c and \ref{AnisoCorr}d), which differs from all other quenches studied.

The squeezing angle dynamics shown in FIG. \ref{Anisoalpha}a and \ref{Anisoalpha}b are similar to the quenches between Ising phases, where the squeezing vector rotates in the clockwise direction at the beginning and the decrease of the squeezing angles $\alpha_s$ is the most rapid at DQPT. In between the first and the second critical times, the squeezing vector turns its rotation direction and the increase in $\alpha_s$ is the fastest at the second critical time. On the other hand, the squeezing angle evolution for quenches from anisotropic boundary (FIG. \ref{Anisoalpha}c and \ref{Anisoalpha}d) reaches its minimum at DQPT, marking the distinction from the cross-boundary quench cases. Then the whole evolution pattern follows what backward quenches have in Ising phases, where it oscillates around $\pi/2$ or $\pi$ and never turns a full round.

\subsection{Quenches from multicritical point}
\label{sec:multcrit}

\begin{figure} [t!]
	\centering
	\includegraphics[width=8.5cm]{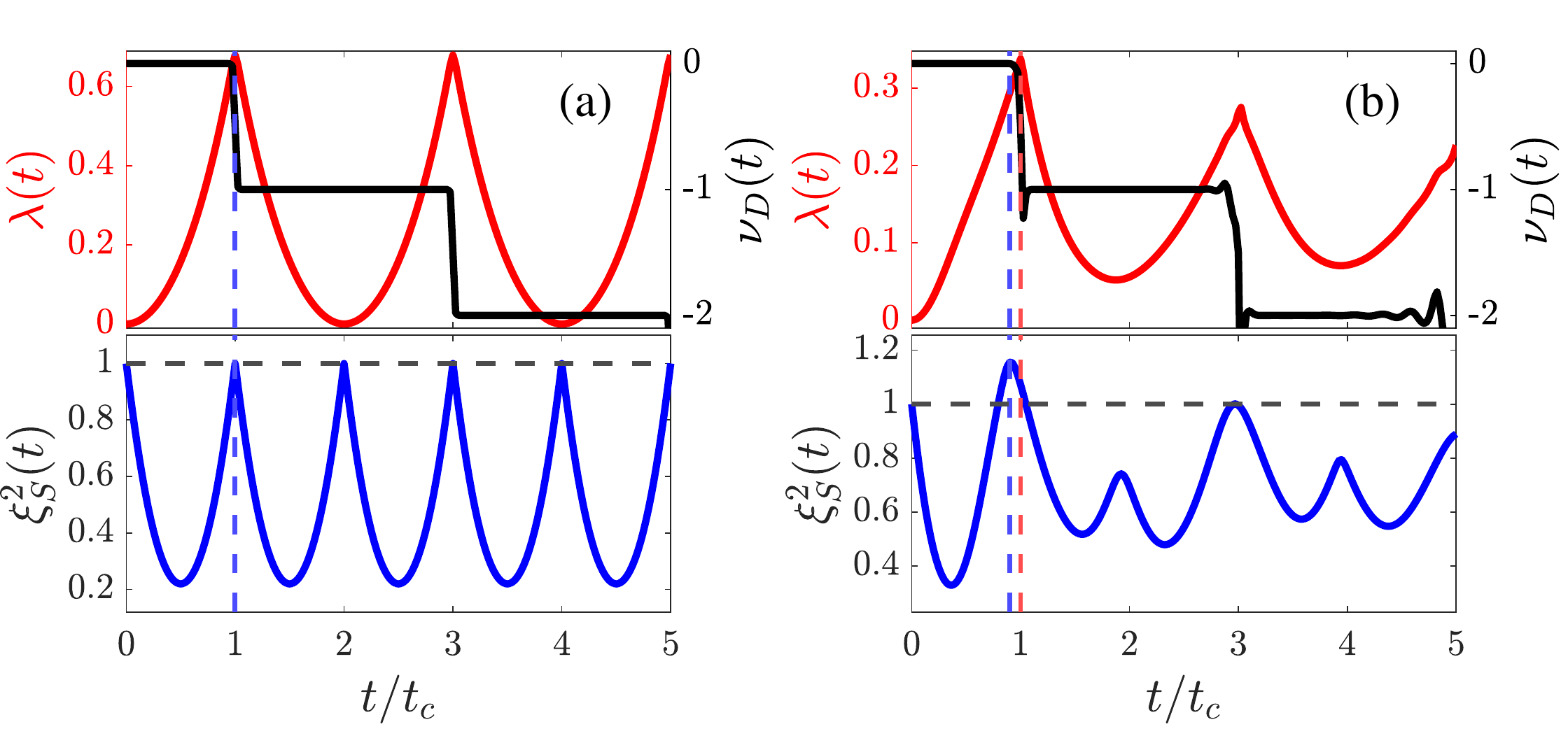}
	\caption{Scaled time plots of LR, DTOP (top panels) and SSP (bottom panels) for (a) $(\delta_i, g_i)=(0,1) \rightarrow (\delta_f, g_f)=(1,0)$ and (b) $(\delta_i, g_i)=(0,1) \rightarrow (\delta_f, g_f)=(0.5,0.5)$. Red and blue dashed lines indicate the first peak of LRs and SSPs respectively and the two lines overlap with each other in (a). Black dashed horizontal lines indicate $\xi_S^2(t) = 1$.}
	\label{MultcritLRSSP}
\end{figure}

\begin{figure} [t!]
	\centering
	\includegraphics[width=8.5cm]{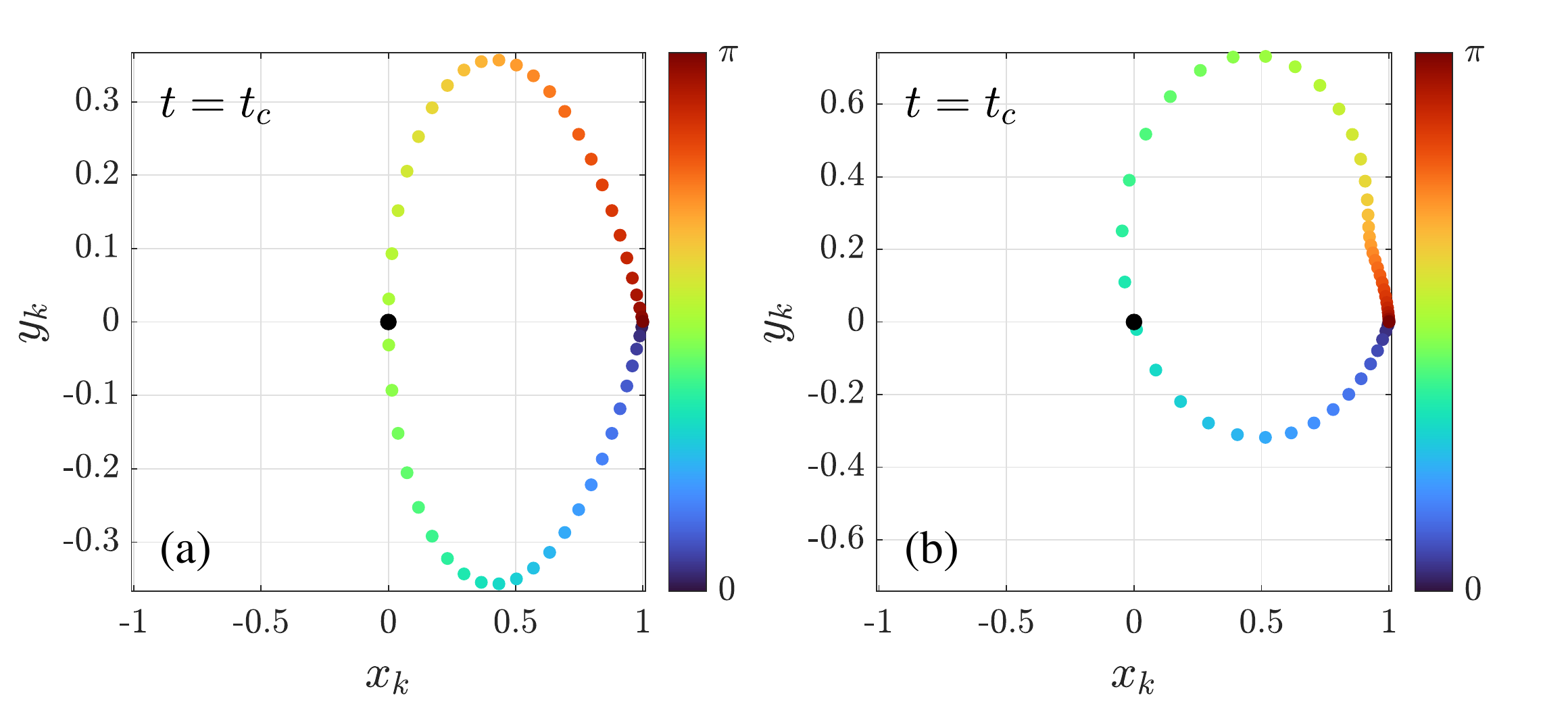}
	\caption{Trajectories of $\vec{r}_k$'s at the corresponding critical times for the same quench cases in FIG. \ref{MultcritLRSSP}. Black dots indicate the origin the vectors are located. The color scale shows the progression of $\vec{r}_k$'s from $k = 0$ to $\pi$.}
	\label{Multcritrk}
\end{figure}

\begin{figure} [t!]
	\centering
	\includegraphics[width=8.5cm]{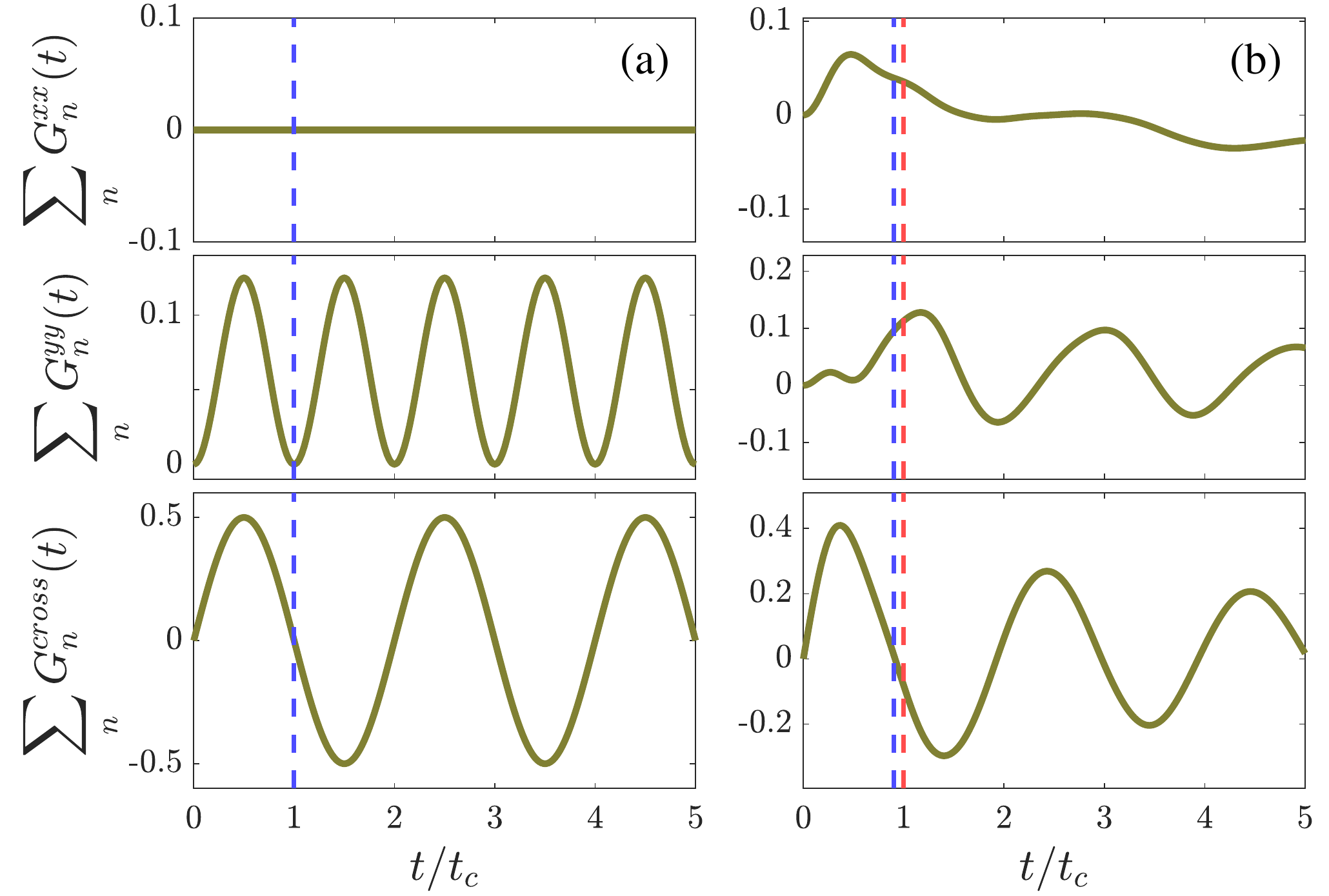}
	\caption{Scaled time plots of sums of correlation functions $G_n^{xx}(t)$ (top panels), $G_n^{yy}(t)$ (middle panels) and $G_n^{cross}(t)$ (bottom panels) for the same quench cases in FIG. \ref{MultcritLRSSP}. Red and blue dashed lines indicate the first peak of LRs and SSPs respectively and the two lines overlap with each other in (a).}
	\label{MultcritCorr}
\end{figure}

\begin{figure} [t!]
	\centering
	\includegraphics[width=8.5cm]{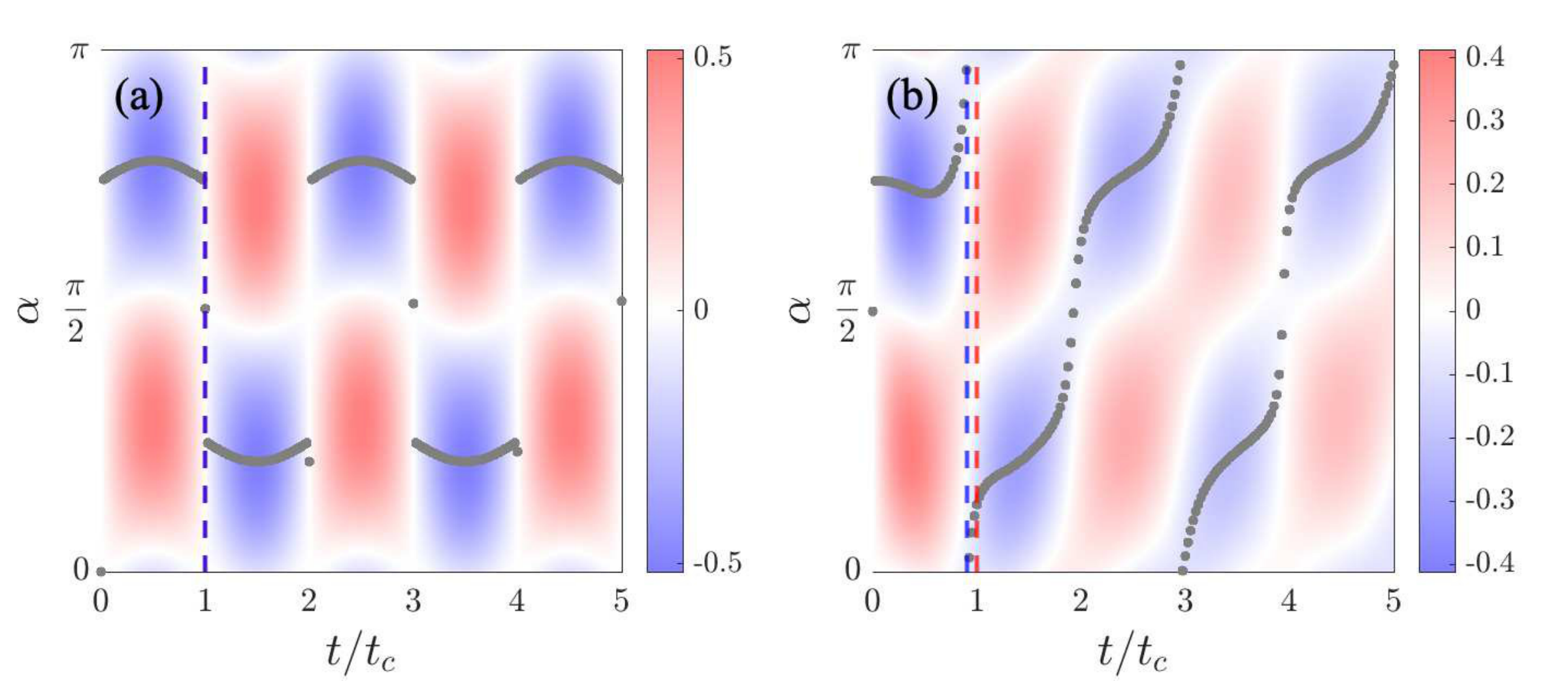}
	\caption{Colormap of $J(\alpha,t)$ for the same quench cases in FIG. \ref{MultcritLRSSP}. Grey dots emphasize the spin squeezing angle $\alpha_s$. Red and blue dashed lines refers to the first peak of LRs and SSPs respectively.}
	\label{Multcritalpha}
\end{figure}

We study two cases when quenched from the multicritical point $(\delta_i,g_i)=(0,1)$. %One involves the end point as the classical Ising model where $(\delta_f,g_f) = (1,0)$ with some exact analytical results. 
Figure \ref{MultcritLRSSP} shows the Loschmidt rate and spin-squeezing parameter evolutions. Notice the perfect oscillation of both LR and SSP for quench to $(\delta_f,g_f) = (1,0)$, the case of a classical Ising model, in FIG. \ref{MultcritLRSSP}a. The periodicity of SSP is half the periodicity of LR. This can be understood from the exact expressions of the LR and SSP, namely the LR is of the form
\begin{equation} \label{0110LR}
	\lambda(t) = \frac{1}{2}\ln8 - \frac{1}{4}\ln[ 38 + 24\cos( 2t ) + 2\cos( 4t ) ]
\end{equation}
and the SSP
\begin{equation} \label{0110SSP}
	\xi_S^2(t) = \frac{1}{8}\bigg[ 9 - \cos( 4t ) - \frac{1}{\sqrt{2}}\sqrt{\cos( 8t ) - 68\cos( 4t ) + 67} \bigg].
\end{equation}
The half periodicity of SSP comes from the dominating $\cos( 4t )$ term in SSP as compared to the dominating $\cos( 2t )$ term in LR in Eq. (\ref{0110LR}) and (\ref{0110SSP}) respectively.

From equations (\ref{0110LR}), we see that the LR is independent of the system size \textit{N}. The size independence also holds for SSP due to the only few non-zero correlation functions as follows:
\begin{equation} \label{multcritCorr}
	\begin{aligned}
		G_2^{yy}(t) &= G_{N - 2}^{yy}(t) = \frac{1}{16}\sin^2( 2t ) \\
		G_1^{xy}(t) &= G_{N - 1}^{xy}(t) = G_1^{yx}(t) = G_{N - 1}^{yx}(t) = \frac{1}{8}\sin( 2t ),
	\end{aligned}
\end{equation}
and all other correlation functions vanish. In other words, in this quench case, all parallel-spin correlations along \textit{x} are zero, whereas the next-nearest neighbor correlations along \textit{y} are finite, resulting in the summed correlation functions in FIG. \ref{MultcritCorr}a. The observation here is counterintuitive to the fact that the system is quenched to the point where the \textit{y}-direction interaction terms are less involved. Notice however that at DQPT, the $G_n^{yy}(t)$'s drop to zero as expected.%The correlation functions are plotted in FIG. \ref{MultcritCorr}a where the relations (\ref{multcritCorr}) are clearly demonstrated. 

The aforementioned perfect pattern in the LR and SSP evolutions are destroyed when $(\delta_f,g_f)$ differs from the classical Ising model. As shown in FIG. \ref{MultcritLRSSP}b, the SSP evolution is distorted and resembles the pattern for a backward quench between Ising phases described in Sec. \ref{sec:Ising}. It has twice as many peaks as the LR, and the major peaks occur near the respective critical times. Nevertheless, for both quench scenarios here, although they start from a critical point, the DTOPs show integer jumps, aligning with the case of quenches across the phase boundary instead of that from a phase boundaries presented in the previous sections.  The trajectories of the vectors $\vec{r}_k(t)$ are also similar to the quenches across the phase boundary (FIG. \ref{Multcritrk}).

The sums of the time evolution of the correlation functions for the quench to $(\delta_f,g_f)$ that differs from the classical Ising model %follow the patterns described in Sec. \ref{sec:Ising} and \ref{sec:aniso}. 
are shown in FIG. \ref{MultcritCorr}b. The distortion emerges and the \textit{x}-direction correlations rise a brief moment and then slowly decrease around DQPT. The perfect oscillation pattern of \textit{yy}-correlation functions is deformed and interestingly they surpass the \textit{xx}-correlation functions at the first critical time. On the other hand, the contribution of cross-spin correlations mostly vanish at DQPT, while before and after DQPT they are of different signs, which resembles the quenches across critical boundary to the FM$_x$ phase.

In terms of the squeezing angle, the quench $(0,1) \rightarrow (1,0)$ acts like the limiting case of backward quenches between Ising phases shown in FIG. \ref{Isingalpha}, where in FIG. \ref{Multcritalpha}a the slope of the squeezing angle approaches $-\infty$ at the first DQPT. The squeezing vector never completes a full rotation, rather it zigzags between the fourth and the first quadrants, i.e. $\alpha_s \in [3\pi / 4,\pi)$ in $0 < t < t_c$ and $\alpha_s \in (0,\pi / 4]$ in $t_c < t < 2t_c$. Note that in the evolution of $\alpha_s$ for the case shown in FIG.\ref{Multcritalpha}b, the squeezing vector rotates anticlockwise and complete a full circle for one revival time. Again we observe the same phenomenon that the change of $\alpha_s$ is the maximum at DQPT.

\section{Discussions}

\begin{figure} [t!]
	\centering
	\includegraphics[width=8.5cm]{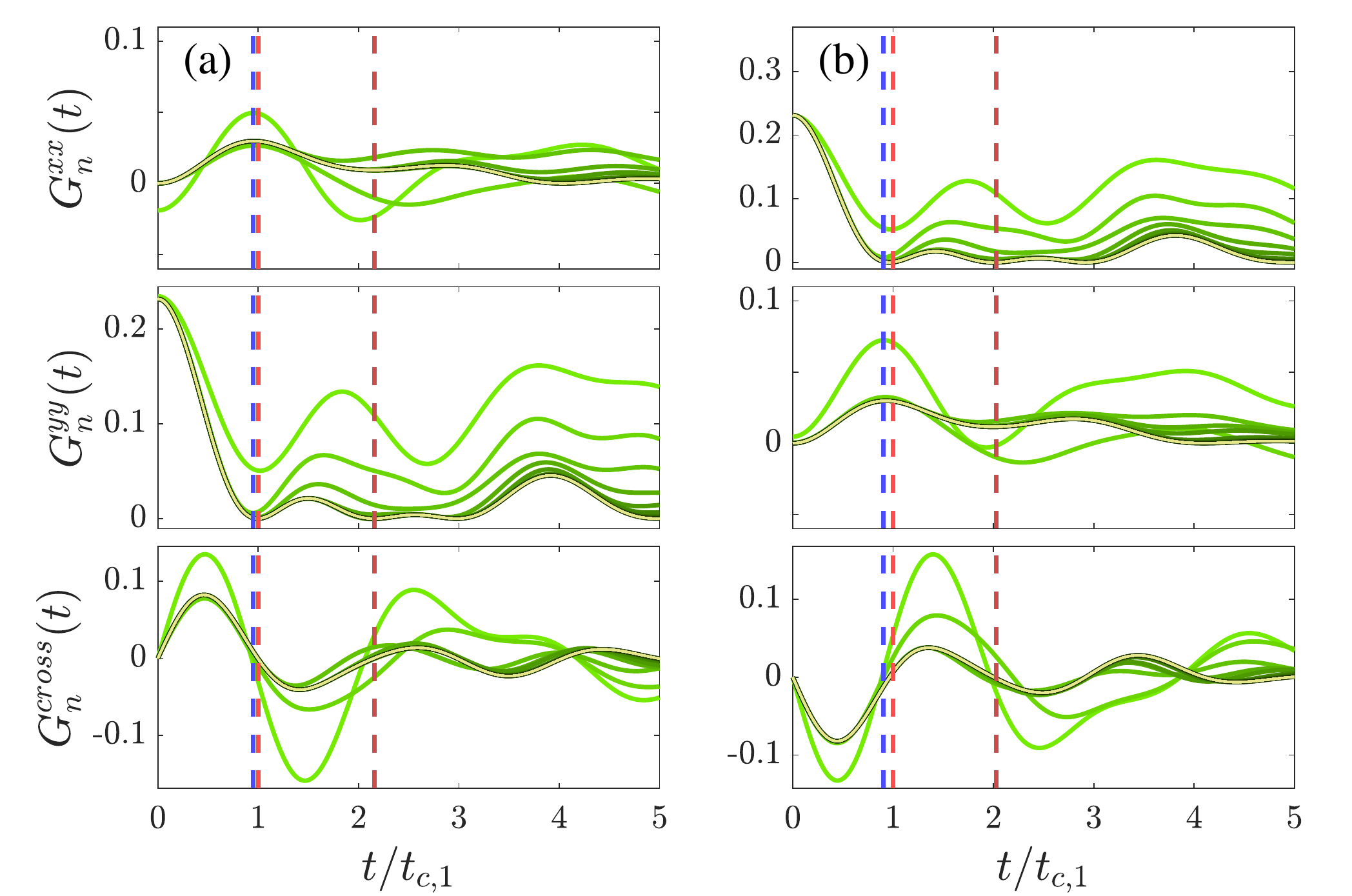}
	\caption{Correlation functions for quenches across anisotropic boundary: (a) $(-1.2,0.5) \rightarrow (0.8,0.5)$ and (b) $(0.8,0.5) \rightarrow (-0.8,0.5)$. We show here $n \leq 10$ from light green to dark green with increasing \textit{n}. Correlations for larger \textit{n}'s converge to the curve in yellow. The first and the second red dashed lines indicate $t_{c,1}$ and $t_{c,2}$ respectively. Blue dashed lines represents the maximum of SSPs.}
	\label{AnisoCorrn}
\end{figure}

\begin{figure} [t!]
	\centering
	\includegraphics[width=8.5cm]{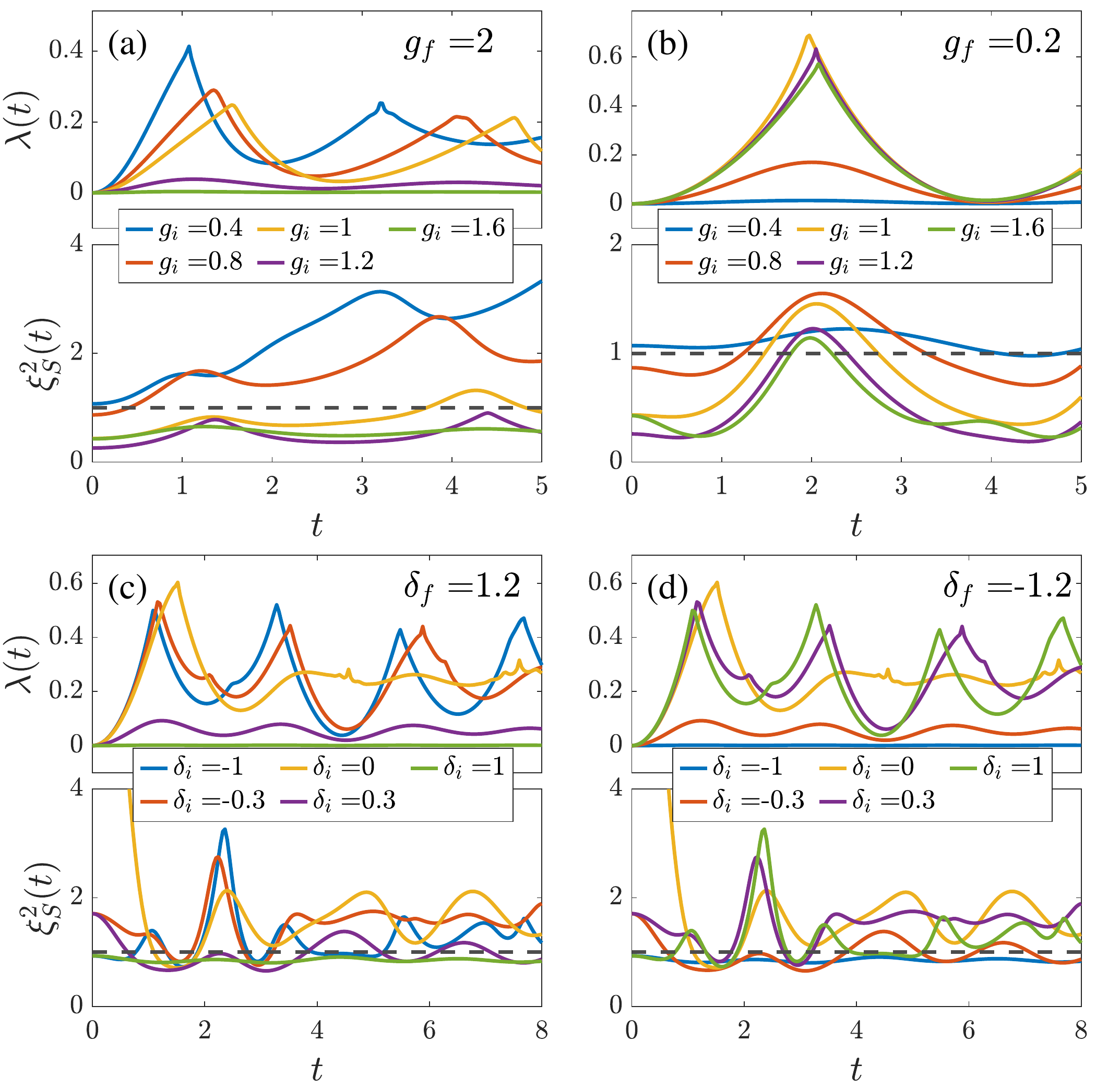}
	\caption{LR and SSP plots of (a),(b) quenches across Ising boundary for fixed $g_f$ with $\delta = 0.8$ and (c),(d) quenches across anisotropic boundary with fixed $\delta_f$ with $g = 0.5$. Black dashed lines indicate the line where $\xi_S^2(t) = 1$.}
	\label{vargideli}
\end{figure}

In the previous section, we showed that the time evolution of the correlation functions matches the physical properties of the phase one quenches to in most of the cases. Here we provide an insight for the whole quench dynamics of an interacting spin model: Once the system is being quenched, the cross-direction correlations either begin growing or shrinking for around half the critical time for all kinds of quench. At the same time for quenches between FM$_x$ and FM$_y$ phases, the parallel-spin correlations evolve according to what phase the system is quenched to -- when the final phase flavors \textit{x}-spin, $G_n^{xx}(t)$'s grow, whereas when the final phase flavors \textit{y}-spin, $G_n^{yy}(t)$'s grow. %\textcolor{blue}{(delete) For quenches involving PM phase, interestingly, the \textit{yy}-spin correlations rise and peak just before DQPT.} 
In the vicinity before the first DQPT, the flavoring parallel-spin correlation functions reach their maximal values while the contribution of cross-spin correlations diminish. In this sense, the system does appear to ``transit" to the final phase at DQPT for a brief amount of time. For the case of quenching to PM phase, interestingly, the supposedly low \textit{yy}-spin correlations rise and peak just before DQPT. However, the originally flavored \textit{xx}-spin correlations shrink to near minimum. %Besides, the increasing SSP implies the growing uncertainty along the spins perpendicular to the MSD, 
This serves as an evidence of ``leaving" the original phase. %Notice however that considering the definition of SSP, this does not assure the certainty of transiting to PM phase due to the fact that the growth of one variance \textcolor{blue}{along one spin direction} does not mean the fall of the other. 
On the other hand, the cross-spin correlation functions rarely play a role during DQPT caused by quenching across the critical boundary, but they do contribute for quenches from critical boundaries, especially for the quench from anisotropic boundary, where the cross-spin correlations contribute a big part in minimizing SSP around DQPT.

Following this interpretation, perhaps we can provide crucial evidence suggesting the second DQPT for quenches across the anisotropic boundary is more of a ``partially return transition" to the initial state. Namely, FIG. \ref{Anisork}(a-d) show the trajectory of the loops including $k_1^*$ (red loop) and $k_2^*$ (blue loop) progresses in opposite direction. %\textcolor{blue}{%winding direction of the vectors for the blue loop (anticlockwise) is opposite to that for the red loop (clockwise)}. 
The strongest evidence comes from the sums of correlation functions in FIG. \ref{AnisoCorr}a and \ref{AnisoCorr}b. In the summed $G_n^{xx}(t)$ and summed $G_n^{yy}(t)$ plots, the first DQPTs occur when the corresponding flavoring correlations of the postquenched Hamiltonian peak, while at the second critical time $t_{c,2}$ this correlations become the lowest. On the contrary, the originally flavored correlations do not restore their domination, yet the nearest-neighbor as well as the first few \textit{n}'s correlations do rise and peak before the second DQPT, as shown in FIG. \ref{AnisoCorrn}, making the summed quantities rise a little between the two critical times. The cross-spin correlation functions vanish as expected for DQPTs. The evolution of the squeezing angles also strengthen the argument, where the squeezing vectors change their rotation directions after $t_{c,1}$ and the squeezing angles pass through $t_{c,2}$ with the greatest slope. After $t_{c,2}$, the squeezing vectors return back to the original rotation direction to complete a full turn. In short, the above evidences suggest the system may partially and temporarily returned to a state having the spin character of the initial state in the vicinity of the second DQPT.

In studying the dynamics around DQPT, often the time-evolved SSP peaks around the critical time. This implies the system is the least squeezed around DQPT. Here we exploit some possible reasons behind regarding the first critical time. From the analytical expression of the SSP in Eq. (\ref{SSPA}), peaked SSP occurs when $G_n^{xx}(t) + G_n^{yy}(t)$ is high, $G_n^{xx}(t) - G_n^{yy}(t)$ and $G_n^{cross}(t)$ are minimal. We can immediately see that for all quench cases we studied, except quenches from phase boundary to the ordered phase, the cross-spin correlation functions are near zero, and the difference of \textit{x} correlations and \textit{y} correlations is minimized. This gives a small square-root term in Eq. (\ref{SSPA}) around DQPT and thus a relatively high SSP. In some cases like the backward quenches in Ising phases and quenches from multicritical point, some $G_n^{yy}(t)$'s are negative in the transient regime, plus the growing contribution from $G_n^{cross}(t)$, the SSP in these cases drops at the beginning. Shortly after, the quantities get lowered to minimal and thus SSPs grow to their maxima near critical time. To put it more generally, when quenched from some phase to the FM$_x$ phase, the sums of $G_n^{xx}(t)$'s are around their maxima and sums of $G_n^{xx}(t)$'s would behave such that $G_n^{xx}(t) - G_n^{yy}(t)$ is low at DQPT. The opposite is true for quenches to FM$_y$ phase. In other words, the \textit{x} and \textit{y} spins are of the most uncertain around DQPT. The cross-spin correlations behave in such a way that vanish at DQPT while before and after DQPT they evolve in opposite sign, making the SSP away from critical time drops.

Besides the different evolutions of spin-related quantities, the overall spin dynamics relating to the types of quenches within one quench scenario is also studied and we found a general relationship between cases with and without successful DQPTs as well as the critical spin dynamics around the first critical time. In FIG. \ref{vargideli}, we plot the rate functions and spin-squeezing parameters for quenches starting from various points in the equilibrium phase diagram. The case involving Ising phases show an apparent distinction from quenches across critical point and within one phase. Namely, SSPs are general high and growing for forward quenches, while low and oscillating below 1 mostly for backward quenches. Quenches within one phases, on the other hand, evolve below 1 for forward quenches and above 1 for backward quenches. The case when the initial point is at the critical point behaves between the two, where the SSP oscillates around 1. Special features can also be observed during successful DQPTs for quenches across anisotropic boundary. In FIG. \ref{vargideli}c and \ref{vargideli}d, quenches across critical point results a sharp peak in SSP at the two critical times, while those peaks are absent in quenches within one phase where no DQPT occurs. Also for the quenches within the phase, the SSPs evolve around and below 1. The critical quench, on the other hand, differs drastically from other cases, where the SSP falls from infinity to a minimum at DQPT, as stated in Sec. \ref{sec:aniso}. In either case, we further confirm the phenomenon that SSP reaches its local extremum (usually maximum) when approaching DQPT.

\section{Conclusion}
\label{sec:conclusion}
%$\mathscr{G}$

We analyzed three distinct quench scenarios in the XY model, for which we successfully detected some physical properties an integrable spin-$\frac{1}{2}$ model might have when quenched between different phases. Firstly, the collective spin of the system tends to fluctuate more when approaching critical times. This can be observed by the time evolution of the spin-squeezing parameter of the system, namely the SSP peaks, or the system is the least spin-squeezed, just before DQPT occurs. This phenomenon seems to persist for longer critical times and for most types of quenches - whether quenched across or from critical boundaries, except when the system is initially in the most unsqueezed state where the SSP diverges. %which again serves as the other extreme spin dynamics. 
In either case, the system exhibits extreme spin-squeezing dynamics relative to the initial state, being the least or the most spin-squeezed situation, in the vicinity of DQPTs.

%SSP itself is also able to distinguish dynamics for different quench types, therefore further investigation of the SSP evolution on other fermionic spin, potentially nonintegrable, models would be appreciated that one might conclude if SSP can be served as a tool to access further spin dynamics of other fermion models and observe similar phenomenon. Possible experiments regarding observing SSP are also encouraged.

By observing the exact analytical expressions for the time evolution of spin-squeezing parameter, we also reveal the spin dynamics by studying the competition between parallel- and cross-spin correlations. We find that during DQPTs, the system does appear to evolve in such a way to match the spin properties of the phase one quenches to. For instance, when quenched to \textit{xx}-spin-flavored phase, the \textit{x}-direction correlations grows to its maximum at DQPTs. This maximum reflects the system is in the least unsqueezed state during DQPTs. Furthermore, we also showed that there is possibly some kind of ``return transition" occurring for models having two critical momenta during the second DQPTs. %The combined observation suggested that there might be some solid phase transition in the dynamic regime of a spin model. 

In short, the time evolution of SSP provides us non-trivial physical insights into the spin correlations around DQPTs, though we would like to remark that the SSP is not a precise order parameter of DQPTs and this work is not intended to address if the spin correlations are the driver of DQPTs. Nevertheless, the above-mentioned observations welcomes further confirmation in experiments, and in other models , for example, the long-range transverse-field Ising model and the Lipkin-Meshhov-Glick model which have been extensively studied for their intriguing dynamical behaviors around various types of phase transitions \cite{Halimeh2017,Homrighausen2017,Corps2022,Corps2023}. It will be interesting to understand the role of long-range interactions in the spin correlation dynamics and the relation to DQPTs. Moreover, the period halving in the SSP peaks relative to the LR observed in some cases also serves along as an interesting future work on how the quenched phases might affect the evolution of spin correlations.

%Lastly, we would like to remark that the SSP is not a precise order parameter of DQPTs and this work is not intended to address if the spin correlations are the driver of DQPTs.   
		
%We believe the potential studies can further confirm whether the demonstrated phenomena} can be proved to be \textcolor{red}{a valid phase transition or even a new type of phase transition}, and thus opening a new phase of understanding the critical dynamics of quantum many-body systems.

\begin{acknowledgments}
We thank Saeed Mahdavifar for the helpful comments on improving the manuscript. This work is financially supported by Research Grants Council  of  Hong  Kong  (Grant  No.  CityU 21304020),  City University of Hong Kong (Grant No. 9610438, 9680320), and the National Science Centre (NCN, Poland) (Grant No. 2019/35/B/ST3/03625).
\end{acknowledgments}


\begin{thebibliography}{}


    % Definitions of DQPT
    
    \bibitem{Heyl2013}
    M. Heyl, A. Polkovnikov, and S. Kehrein, Phys. Rev. Lett. \textbf{110}, 135704 (2013).
    
    \bibitem{Heyl2018}
    M. Heyl, Rep. Prog. Phys. \textbf{81}, 054001 (2018).
    
    \bibitem{Zvyagin2016}
    A. A. Zvyagin, Low Temp. Phys. \textbf{42}, 971 (2016).
    
    %Type-I DQPT
    \bibitem{Zunkovic2018}
    B. \v{Z}unkovi\v{c}, M. Heyl, M. Knap, and A. Silva, Phys. Rev. Lett. \textbf{120}, 130601 (2018).

     \bibitem{Halimeh2017}
    J. C. Halimeh, and V. Zauner-Stauber, Phys. Rev. B \textbf{96}, 134427 (2017).
    
    \bibitem{Yuzbashyan2006}
    E. A. Yuzbashyan, O. Tsyplyatyev, and B. L. Altshuler, Phys. Rev. Lett. \textbf{96}, 097005 (2006).
    
    \bibitem{Sciolla2010}
    B. Sciolla, and G. Biroli, Phys. Rev. Lett. \textbf{105}, 220401 (2010).
    
    \bibitem{Zunkovic2016}
    B. \v{Z}unkovi\v{c}, A. Silva, and M. Fabrizio, Phil. Trans. R. Soc. A \textbf{374}, 20150160 (2016).
    %
    
    % Experimental works on DQPT
        
    \bibitem{Tian2019}
    T. Tian, Y. Ke, L. Zhang, S. Lin, Z. Shi, P. Huang, C. Lee, and J. Du, Phys. Rev. B \textbf{100}, 024310 (2019).
    
    \bibitem{Jurcevic2017}
    P. Jurcevic, H. Shen, P. Hauke, C. Maier, T. Brydges, C. Hempel, B. P. Lanyon, M. Heyl, R. Blatt, and C. F. Roos, Phys. Rev. Lett. \textbf{119}, 080501 (2017).
    
    \bibitem{Zhang2017}
    J. Zhang, G. Pagano, P. W. Hess, A. Kyprianidis, P. becker, H. Kaplan, A. V. Gorshkov, Z. -X. Gong, and C. Monroe, Nature (London) \textbf{511}, 601 (2017).
    
        \bibitem{Flaschner2018}
    N. Fl\"{a}schner, D. Vogel, M. Tarnowski, B. S. Rem, D.-S. L\"{u}hmann, M. Heyl, J. C. Budich, L. Mathey, K. Sengstock, and C. Weitenberg, Nature Phys \textbf{14}, 265 (2018).
    
        \bibitem{Guo2019}
    X. -Y. Guo, C. Yang, Y. Zeng, Y. Peng, H. -K. Li, H. Deng, Y. -R. Jin, Sh. Chen, D. Zheng, and H. Fan, Phys. Rev. Applied \textbf{11}, 044080 (2019).
    
    \bibitem{Yang2019}
    H.-X. Yang, T. Tian, Y.-B. Yang, L.-Y. Qiu, H.-Y. Liang, A.-J. Chu, C. B. Da\u{g}, Y. Xu, Y. Liu, and L.-M. Duan, Phys. Rev. A \textbf{100}, 013622 (2019).
    
    \bibitem{Tian2020}
    T. Tian, H.-X. Yang, L.-Y. Qiu, H.-Y. Liang, Y.-B. Yang, Y. Xu, and L.-M. Duan, Phys. Rev. Lett. \textbf{124}, 043001 (2020).
    
    \bibitem{Xu2020}
    K. Xu, Z.-H. Sun, W. Liu, Y.-R. Zhang, H. Li, H. Dong, W. Ren, P. Zhang, F. Nori, D. Zheng, H. Fan, and H. Wang, Sci. Adv. (2020) \textbf{6}: eaba4935.
    
    \bibitem{Bernien2017}
    H. Bernien, S. Schwartz, A. Keesling, H. Levine, A. Omran, H. Pichler, S. Choi, A. S. Zibrov, M. Endres, M. Greiner, V. Vuleti\'{c}, and M. D. Lukin, Nature (London) \textbf{551}, 579 (2017).
    
    \bibitem{Sanchez2018}
    E. Guardado-Sanchez, P. T. Brown, D. Mitra, T. Devakul, D. A. Huse, P. Schauss, and W. S. Bakr, Phys. Rev. X \textbf{8}, 021069 (2018).
    
    %Theoretical works on DQPT
%    \bibitem{DoLEEbQC}
%    H. T. Quan, Z. Song, X. F. Liu, P. Zanardi, and C. P. Sun, Phys. Rev. Lett. \textbf{96}, 140604 (2006).
    
    \bibitem{Heyl2015}
    M. Heyl, Phys. Rev. Lett. \textbf{115}, 140602 (2015).
    
    \bibitem{Heyl2014}
    M. Heyl, Phys. Rev. Lett. \textbf{113}, 205701 (2014).
    
    \bibitem{Weidinger2017}
    S. A. Weidinger, M. Heyl, A. Silva, and M. Knap, Phys. Rev. B \textbf{96}, 134313 (2017).
    
    \bibitem{Titum2019}
    P. Titum, J. T. Iosue, J. R. Garrison, A. V. Gorshkov, and Z. -X. Gong, Phys. Rev. Lett. \textbf{123}, 115701 (2019).
    
    \bibitem{Jafari2019}
    R. Jafari, Sci. Rep. \textbf{9}, 2871 (2019).
    
    \bibitem{Schmitt2015}
    M. Schmitt, and S. Kehrein, Phys. Rev, B \textbf{92}, 075114 (2015).
    
    \bibitem{Schmitt2018}
    M. Schmitt, and M. Heyl, SciPost Phys. \textbf{4}, 013 (2018).
    
    \bibitem{Torlai2014}
    G. Torlai, L. Tagliacozzo, and G. D. Chiara, J. Stat. Mech. (2014) P06001.
    
    \bibitem{Canovi2014}
    E. Canovi, E. Ercolessi, P. Naldesi, L. Taddia, and D. Vodola, Phys. Rev. B \textbf{89}, 104303 (2014).
    
    \bibitem{Sedlmayr2018}
    N. Sedlmayr, P. Jaeger, M. Maiti, and J. Sirker, Phys. Rev. B \textbf{97}, 064304 (2018).
    
    \bibitem{Poyhonen2021}
    K. P\"{o}yh\"{o}nen and T. Ojanen, Phys. Rev. Res. \textbf{3}, L042027 (2021).
    
    \bibitem{Vosk2014}
    R. Vosk and E. Altman, Phys. Rev. Lett. \textbf{112}, 217204 (2014).
    
    \bibitem{Patra2011}
    A. Patra, V. Mukherjee, and A. Dutta, J. Phys. Conf. Ser. \textbf{297}, 012008 (2011).
    
  %  \bibitem{QQitRFIC}
   % P. Calabrese, F. H. L. Essler, and M. Fagotti, Phys. Rev. Lett. \textbf{106}, 227203 (2011).
    
    \bibitem{Nicola2020}
    S. De Nicola, B. Doyon, and M. J. Bhaseen, J. Stat. Mech. (2020) 013106.
    
    \bibitem{Nicola2021}
    S. De Nicola, A. A. Michailidis, and M. Serbyn, Phys. Rev. Lett. \textbf{126}, 040602 (2021).
    
    \bibitem{HeylPollmann2018}
    M. Heyl, F. Pollmann, and B. D\'{o}ra, Phys. Rev Lett. \textbf{121}, 016801 (2018).
    
    
  %  \bibitem{DaqiodqIls}
  %  D. Rossini, and E. Vicari, Phy. Rev. B \textbf{103}, 179901 (2021).
    
    \bibitem{Budich2016}
    J. C. Budich, and M. Heyl, Phys. Rev. B \textbf{93}, 085416 (2016).
    
    \bibitem{Yu2021}
    W. C. Yu, P. D. Sacramento, Y. C. Li, and H. -Q. Lin, Phys. Rev. B \textbf{104}, 085104 (2021).
    
    \bibitem{Zache2019}
    T. V. Zache, N. Mueller, J. T. Schneider, F. Jendrzejewski, J. Berges, and P. Hauke, Phys. Rev. Lett. \textbf{122}, 050403 (2019).
    
    \bibitem{Vajna2014}
    S. Vajna, and B. D\'{o}ra, Phys. Rev. B \textbf{89}, 161105(R) (2014).
    
    \bibitem{Lacki2019}
    M. Lacki, and M. Heyl, Phys. Rev. B \textbf{99}, 121107(R) (2019).
    
    \bibitem{Lahiri2019}
    A. Lahiri, and S. Bera, Phys. Rev. B \textbf{99}, 174311 (2019).
    
    \bibitem{YangZhou2019}
    K. Yang, L. Zhou, W. Ma, X. Kong, P. Wang, X. Qin, X. Rong, Y. Wang, F. Shi, J. Gong, and J. Du, Phys. Rev. B \textbf{100}, 085308 (2019).
    
    \bibitem{Kosior2018}
    A. Kosior, and K. Sacha, Phys. Rev. A \textbf{97}, 053621 (2018).
    
    \bibitem{Kyaw2020}
    T. H. Kyaw, V. M. Bastidas, J. Tangpanitanon, G. Romero, and L. C. Kwek, Phys. Rev. A \textbf{101}, 012111 (2020).
    
    \bibitem{Zvyagin2017}
    A. A. Zvyagin, Phys. Rev. B \textbf{95}, 075122 (2017).
    
    \bibitem{JafariJohannesson2019}
    R. Jafari, H. Johannesson, A. Langari, and M. A. Martin-Delgado, Phys. Rev. B \textbf{99}, 054302 (2019).
    
    \bibitem{Jafari2021}
    R. Jafari, and A. Akbari, Phys. Rev. A \textbf{103}, 012204 (2021).
    
    \bibitem{Zamani2020}
    S. Zamani, R. Jafari, and A. Langari, Phys. Rev. B \textbf{102}, 144306 (2020).
    
    \bibitem{Bandyopadhyay2021}
    S. Bandyopadhyay, A. Polkovnikov, and A. Dutta, Phys. Rev. Lett. \textbf{126}, 200602 (2021).
    
    \bibitem{Halimeh2021}
    J. C. Halimeh, D. Trapin, M. VanDamme, and M. Heyl, Phys. Rev. B \textbf{104}, 075130 (2021).
    
    \bibitem{Markov2021}
    A. A. Markov, and A. N. Rubtsov, Phys. Rev. B \textbf{104}, L081105 (2021).
    
    \bibitem{Homrighausen2017}
    I. Homrighausen, N. O. Abeling, V. Zauner-Stauber, and J. C. Halimeh, Phys. Rev. B \textbf{96}, 104436 (2017).
	
	%TFIM
	

	
	\bibitem{Ding2020}
	C. Ding, Phys. Rev. B \textbf{102}, 060409(R) (2020).
	
	%ANNNI model
	\bibitem{Karrasch2013}
	C. Karrasch, and D. Schuricht, Phys. Rev. B \textbf{87}, 195104 (2013).
	
	\bibitem{Kriel2014}
	J. N. Kriel, C. Karrasch, and S. Kehrein, Phys. Rev. B \textbf{90}, 125106 (2014).
	
	\bibitem{Cheraghi2020}
	H. Cheraghi, M. J. Tafreshi, and S. Mahdavifar, J. Magn. Magn. Mater. \textbf{497}, 166078 (2020).
	
	\bibitem{Kennes2018}
	D. M. Kennes, D. Schuricht, and C. Karrasch, Phys. Rev. B \textbf{97}, 184302 (2018).
	
	
	\bibitem{Seetharam2021}
	K. Seetharam, Y. Shchadilova, F. Grusdt, M. B. Zvonarev, and E. Demler, Phys. Rev. Lett. \textbf{127}, 185302 (2021).
	
	\bibitem{Sadrzadeh2021}
	M. Sadrzadeh, R. Jafari, and A. Langari, Phys. Rev. B \textbf{103}, 144305 (2021).
	
	\bibitem{Naji2022}
	J. Naji, M. Jafari, R. Jafari, and A. Akbari, Phys. Rev. A \textbf{105}, 022220 (2022).
	
	\bibitem{Jafari2022}
	R. Jafari, A. Akbari, U. Mishra, and H. Johannesson, Phys. Rev. B \textbf{105}, 094311 (2022).
	
	\bibitem{Mishra2020}
	U. Mishra, R. Jafari, and A. Akbari, J. Phys. A: Math. Theor. \textbf{53}, 375301 (2020).
	
	\bibitem{Halimeh2020}
	J. C. Halimeh, M. Van Damme, V. Zauner-Stauber, and L. Vanderstraeten, Phys. Rev. Res. \textbf{2}, 033111 (2020).
	
	\bibitem{Hashizume2022}
	T. Hashizume, Ian P. McCulloch, and Jad C. Halimeh, Phys. Rev. Res. \textbf{4}, 013250 (2022).
	
	\bibitem{Damme2022}
	M. Van Damme, T. V. Zache, D. Banerjee, P. Hauke, and J. C. Halimeh, Phys. Rev. B \textbf{106}, 245110 (2022).
	
	\bibitem{Uhrich2020}
	P. Uhrich, N. Defenu, R. Jafari, and J. C. Halimeh, Phys. Rev. B \textbf{101}, 245148 (2020).
	
	\bibitem{Lang2018}
	J. Lang, B. Frank, and J. C. Halimeh, Phys. Rev. Letts. \textbf{121}, 130603 (2018).
	
	\bibitem{Andraschko2014}
	F. Andraschko and J. Sirker, Phys. Rev. B \textbf{89}, 125120 (2014).
	
	\bibitem{Morrison2008}
	S. Morrison and A. S. Parkins, Phys. Rev. Lett. \textbf{100}, 040403 (2008).
	
	\bibitem{Porta2020}
	S. Porta, F. Cavaliere, M. Sassetti, and N. T. Ziani, Sci. Rep. \textbf{10}, 12766 (2020).
	
	\bibitem{Wong2022}
	C. Y. Wong and W. C. Yu, Phys. Rev. B \textbf{105}, 174307 (2022).
	
	\bibitem{Corps2022}
	\`{A}. L. Corps and A. Rela\~{n}o, Phys. Rev. B \textbf{106}, 024311 (2022).
	
	\bibitem{Corps2023}
	\`{A}. L. Corps and A. Rela\~{n}o, Phys. Rev. Lett. \textbf{130}, 100402 (2023).
	
    
    %SSP

    
    \bibitem{Ma2011}
    J. Ma, X. Wang, C. P. Sun, and F. Nori, Phys. Rep. \textbf{509}, 89 (2011).
    
    \bibitem{Gross2012}
    C. Gross, J. Phys. B: At. Mol. Opt. Phys. \textbf{45}, 103001 (2012).
    
    \bibitem{Pezze2018}
    L. Pezz\`{e}, A. Smerzi, M. K. Oberthaler, R. Schmied, and P. Treutlein, Rev. Mod. Phys. \textbf{90}, 035005 (2018).
    
    \bibitem{Hamley2012}
    C. D. Hamley, C. S. Gerving, T. M. Hoang, E. M. Bookjans, and M. S. Chapman, Nat. Phys. \textbf{8}, 305 (2012).
    
    \bibitem{Sun2011}
    Z. Sun, Phys. Rev. A \textbf{84}, 052307 (2011).
    
    \bibitem{Vidal2004}
    J. Vidal, G. Palacios, and R. Mosseri, Phys. Rev. A \textbf{69}, 022107 (2004).
    
    \bibitem{Huang2022}
    Y. Huang, Y. Ding, J. Xu, J. Liu, H. Wang, and H. -N. Xiong, Phys. Rev. A \textbf{106}, 022430 (2022).
    
    \bibitem{Cheraghi2022}
    H. Cheraghi, S. Mahdavifar, and H. Johannesson, Phys. Rev. B \textbf{105}, 024425 (2022).
    
    \bibitem{Kitagawa1993}
    M. Kitagawa and M. Ueda, Phys. Rev. A \textbf{47}, 5138 (1993).
    
    %Analytical XY
    
    \bibitem{Sachdev}
    S. Sachdev, \textit{Quantum Phase Transitions} (Cambridge University Press, Cambridge, England, 1999).
    
    \bibitem{Barouch1970}
    E. Barouch, B. M. McCoy, and M. Dresden, Phys. Rev. A \textbf{2}, 1075 (1970).
    
    \bibitem{Barouch1971}
    E. Barouch and B. M. McCoy, Phys. Rev. A \textbf{3}, 786 (1971).
     
     \bibitem{Pancharatnam1956}
     S. Pancharatnam, Proc. Indian Acad. Sci. A \textbf{44}, 247 (1956).
     
     \bibitem{Samuel1988}
     J. Samuel and R. Bhandari, Phys. Rev. Lett. \textbf{60}, 2339 (1988).
     
     %Squeezing angle
     \bibitem{Jin2007}
     G. -R. Jin and S. W. Kim, Phys. Rev. A \textbf{76}, 043621 (2007).
     %
     
     \bibitem{Lieb1961}
     E. Lieb, T. Schultz, and D. Mattis, Ann. Phys. \textbf{16}, 407 (1961).
    
    
    % MBL
       % \bibitem{Nandkishore2015}
    %R. Nandkishore, and D. A. Huse, Annu. Rev. Condens. Matter Phys. 6, 15 (2015).
    
    %\bibitem{Abanin2017}
    %D. A. Abanin, and Z. Papic, Ann. Phys. (Berlin) 529, 1700169 (2017).
    
    
    
    %\bibitem{Sacramento2016}
    %P. D. Sacramento, Phys. Rev. E \textbf{93}, 062117 (2016).

    
    %\bibitem{Yu2016}
	%W. C. Yu, and S. J. Gu, Chin. Phys. B \textbf{25}, 030501 (2016).
	
	%\bibitem{Serbyn2015}
	%M. Serbyn, Z. Papi\'{c}, and D. A. Abanin, Phys. Rev. X \textbf{5}, 041047 (2015).
	
	%\bibitem{Sachdev1999}
	%S. Sachdev, \textit{Quantum Phase Transitions} (Cambridge University Press, Cambridge, 1999).
	
		%\bibitem{SotWDoaQCSbQaCP}
	%A. Silva, Phys. Rev. Lett. \textbf{101}, 120603 (2008).
	
	%\bibitem{SuppMat}
	%For LAS with randomly selected excited states, see the Supplemental Materials.

\end{thebibliography}
\end{document}